\documentclass[twocolumn,showpacs,nofootinbib,preprintnumbers,amsmath,amssymb,showkeys]{revtex4}

\usepackage{epsfig}
\usepackage{dsfont}

\usepackage{graphicx}
\usepackage{dcolumn}
\usepackage{bm}

\newcommand{\beq}{\begin{equation}}
\newcommand{\eeq}{\end{equation}}
\newcommand{\bea}{\begin{eqnarray}}
\newcommand{\eea}{\end{eqnarray}}
\newcommand{\nn}{\nonumber \\}

\newcommand\eqn[1]{(\ref{#1})}      
\newcommand\Eqn[1]{Eq.~(\ref{#1})}  

\newcommand{\C}{\mathcal{C}}

\newcommand{\bx}{{\bf x}}

\newcommand{\bq}{{\bf q}}
\newcommand{\be}{{\bf e}}

\newcommand{\bX}{{\bf X}}
\newcommand{\bK}{{\bf K}}
\newcommand{\bS}{{\bf S}}

\begin{document}


\title{Nonperturbative resummation of de Sitter infrared logarithms\\in the large-$N$ limit}

\author{J. Serreau}%
 \email{serreau@apc.univ-paris7.fr}
\affiliation{%
 Astro-Particule et Cosmologie (APC), CNRS UMR 7164, Universit\'e Paris 7 - Denis Diderot\\ 10, rue Alice Domon et L\'eonie Duquet, 75205 Paris Cedex 13, France
}%
\author{R. Parentani}
 \email{parentani@th.u-psud.fr}
\affiliation{%
 Laboratoire de Physique Th\'eorique (LPT), CNRS UMR 8627, B\^at. 210, Universit\'e Paris - Sud 11, 91405 Orsay Cedex, France
}%

\date{\today}

\begin{abstract}

We study the $O(N)$ scalar field theory with quartic self-coupling in de Sitter space. When the field is light in units of the expansion rate, perturbative methods break down at very low momenta due to large infrared logarithmic terms. Using the nonperturbative large-$N$ limit, we compute the four-point vertex function in the deep infrared regime. The resummation of an infinite series of perturbative (bubble) diagrams leads to a modified power law which is analogous to the generation of  an anomalous dimension in critical phenomena. We discuss in detail the role of high momentum (subhorizon) modes, including the issue of renormalization, and show that they influence the dynamics of infrared (superhorizon) modes only through a constant renormalization factor. This provides an explicit example of effective decoupling between high and low energy physics in an expanding space-time.

 \end{abstract}

\pacs{11.10.-z, 04.62.+v}
\keywords{Quantum field theory, de Sitter space, large-$N$ techniques}
\maketitle


\section{Introduction}
\label{sec:intro}

The study of quantum field dynamics on de Sitter space has received strong phenomenological motivations with the impressive observational success of the inflation paradigm and the need to compute quantum corrections to inflationary observables and with the observation of the recent acceleration of the Universe. Radiative corrections on de Sitter space have been addressed in a variety of field theories, mainly based on perturbative loop expansion \cite{Tsamis:1996qk,Prokopec:2002jn,Brunier:2004sb,Weinberg:2005vy,Sloth:2006az,Seery:2007we,vanderMeulen:2007ah}. In the phenomenologically relevant case of a light field, with a mass small as compared to the Hubble parameter, loop diagrams typically exhibit large infrared (IR) logarithms, unless they are protected by some symmetries \cite{Geshnizjani:2002wp,Urakawa:2009my,Senatore:2009cf,Kahya:2010xh}. In addition, in cases where the de Sitter symmetry is broken, e.g., by finite time initial conditions, loop corrections give secular divergences which grow as powers of the number of e-folds \cite{Starobinsky:1994bd,Onemli:2002hr,Tsamis:2005hd,Weinberg:2005vy,Burgess:2009bs,Kahya:2010xh}. 
The understanding of such IR/secular divergences is of key importance for a fundamental description of quantum field dynamics in de Sitter space. For instance, it is unknown whether they signal an instability of de Sitter space against quantum fluctuations \cite{Ford:1984hs,Mottola:1985qt,Antoniadis:1985pj,Tsamis:1996qq,Polyakov:2009nq,Marolf:2010nz,Hollands:2010pr}. In any case, such divergences pose a problem of principle and call for resummation.

The situation is somewhat similar to the appearance of large logarithms in standard quantum field theory (QFT). These signals a breakdown of perturbation theory and their resummation lead to nontrivial phenomena such as running couplings or anomalous scaling of field correlators \cite{Weinberg:1996kw,Delamotte:2007pf}. IR and/or secular divergences are common features of perturbative approaches e.g. for bosonic theories at high temperatures \cite{Blaizot:2003tw} or near a second order phase transition \cite{Delamotte:2007pf} or in the case of nonequilibrium systems \cite{Berges:2004vw}. 
Various resummation techniques or nonperturbative approaches have been developed over the years to cope with such issues in flat space-time. These include, e.g., large-$N$ techniques \cite{Coleman:1974jh,Root:1974zr}, the renormalization group \cite{Delamotte:2007pf,Boyanovsky:1998aa}, hard thermal loops \cite{Blaizot:2003tw}, or two-particle-irreducible (2PI) techniques \cite{Cornwall:1974vz,Berges:2004vw}. Recently, some effort has been put in trying to extend these techniques to the case of IR/secular divergences in de Sitter space \cite{Riotto:2008mv,Garbrecht:2011gu,Burgess:2009bs,Serreau:2011fu,Prokopec:2011ms,Arai:2011dd,Boyanovsky:2012qs,Rajaraman:2010xd,Beneke:2012kn,Parentani:2012tx,Akhmedov:2011pj,Youssef:2013by}. 

The large-$N$ limit in $O(N)$ scalar theories provides a simple nonperturbative approach which captures nontrivial IR physics in flat space-time \cite{Coleman:1974jh,Root:1974zr,Cooper:1994hr,Aarts:2002dj,Baacke:2004dp}. Recently this approach has been applied in de Sitter space with interesting results \cite{Riotto:2008mv,Serreau:2011fu}. For instance, it describes dynamical mass generation, first pointed out in \cite{Starobinsky:1994bd}, which screens perturbative IR divergences and renders the theory well behaved. The generated mass exhibits a nonanalytic dependence in the coupling, which is typical of nonperturbative IR physics. Also it has been shown that strong IR fluctuations prohibit the possibility of a spontaneously broken phase \cite{Serreau:2011fu,Ratra:1984yq}, a phenomenon akin to what happens in two-dimensional flat space-time \cite{Mermin:1966fe}.

The phenomena described here involve the resummation of a particular class of IR divergences corresponding to purely local contributions to the self-energy. Such masslike contributions are specific to theories with quartic self-interactions and are technically easy to deal with due to their local character. Less simple but more generic are the logarithmic IR divergencies arising from nonlocal perturbative diagrams. 

In the present paper we consider the four-point vertex function of an $O(N)$ scalar field theory. In the large-$N$ limit, the latter is given by an infinite series of nonlocal bubble loop diagrams, each of which bringing additional IR logarithms. Exploiting the physical momentum space representation (hereafter called $p$-representation) for de Sitter correlators \cite{Busch:2012ne,Parentani:2012tx,Adamek:2013vw}, we show that this series can be exactly resummed. This induces a modification of the IR power law behavior in a similar way as the resummation of large logarithms in standard QFT, e.g., at a critical point, produces an anomalous scaling of field correlators. 

The nonlocal character of the physics described above leads to an nontrivial interplay between IR and ultraviolet (UV) momentum modes. In the main body of the paper, we simply disregard high-momentum modes. This allows us to exhibit the anomalous dimension phenomenon in arbitrary dimension. We show that the results are essentially insensitive to the scale which separates IR and UV modes. A detailed analysis of the role of UV modes in four space-time dimension---including a discussion of renormalization---is presented in the Appendices together with some technical material needed for the main body of the paper. Remarkably, we find that UV modes do not alter the IR power law behaviors and merely give rise to a finite computable constant renormalization factor. This demonstrates a form of effective decoupling between IR and UV physics in de Sitter.

In Sec. \ref{sec:prep} we briefly review the main aspects of the $p$-representation of de Sitter correlators. Then Sec. \ref{sec:ON} focuses on the $O(N)$ scalar theory in the large-$N$ limit for both the two- and the four-point vertex functions. We recall the solution of the corresponding equations for the two-point function, which describes self-consistent mass generation in Sec. \ref{sec:2pt}. The analysis of the four-point function and the resummation of nonlocal IR logarithms is performed in Sec. \ref{sec:four}, where we neglect UV modes. The important technical details needed for this analysis are presented in Appendices \ref{appsec:proof}-\ref{appsec:piIR}. Finally, Appendices \ref{appsec:pimix}-\ref{appsec:renorm} present a detailed treatment of UV modes in four space-time dimensions, where the present theory is perturbatively renormalizable.

\section{$p$-representation of two-point correlators in de Sitter space}
\label{sec:prep}

We consider a generic quantum scalar field $\varphi(x)$ on the expanding Poincar\'e patch of de Sitter space-time with $D=d+1$ dimensions and expansion rate $H=1$. The line element is given by
\beq
\label{eq:line}
 ds^2=a^2(\eta)\left(-d\eta^2+d\bX^2\right)\nn
\eeq
in terms of the conformal time $\eta$ and comoving spatial coordinates $\bX$, with $a(\eta)=-1/\eta=e^t$ with $t\in \mathbb{R}$. 

\begin{figure}[t!]  
\epsfig{file=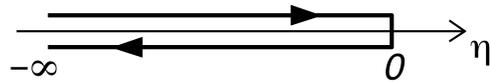,width=6.5cm}
 \caption{\label{fig:etapath} 
The closed path $\C=\C^+\cup\C^-$ in conformal time $\eta$. The forward (upper) branch $\C^+$ goes from $-\infty$ to $0^-$ and the backward (lower) branch $\C^-$ goes back from $0^-$ to $-\infty$.}
\end{figure}

Like general nonequilibrium quantum systems \cite{Schwinger:1960qe,Bakshi:1962dv,Keldysh:1964ud,Chou:1984es,Berges:2004vw}, quantum field theories in cosmological spaces can be conveniently formulated using a closed contour $\C$ in the time coordinate \cite{Ramsey:1997qc,Tranberg:2008ae,Parentani:2012tx}. The appropriate contour for conformal time is depicted in Fig.~\ref{fig:etapath}. The various components of $n$-point correlators are then described by means of time ordered products of field operators on the contour. For instance the two-point function $G(x,x')=\langle T_\C\varphi(x)\varphi(x')\rangle$, where $T_\C$ denotes time ordering on the contour $\C$, encodes both the statistical and spectral correlators\footnote{We use $\{A,B\}=AB+BA$ and $[A,B]=AB-BA$.} $F(x,x')={1\over2}\langle\{\varphi(x),\varphi(x')\}\rangle$ and $\rho(x,x')=i\langle[\varphi(x),\varphi(x')]\rangle$:
\beq
\label{eq:ttt}
 G(x,x')=F(x,x')-\frac{i}{2}{\rm sign}_\C(x^0-x^{\prime0})\rho(x,x'),
\eeq
where the sign function is to be understood along the contour $\C$.
In the rest of this subsection, we consider the statistical function $F$. Everything holds equally for the spectral function $\rho$.

De Sitter invariance ensures that the two-point function $F(x,x')$ only depends on the invariant distance $z(x,x')$. In the coordinate system \eqn{eq:line}, the latter reads
\beq
 z(x,x')=\frac{\eta^2+\eta^{\prime2}-(\bX-\bX')^2}{2\eta\eta'}.
\eeq
The dependence on comoving spatial variable is only through $|\bX-\bX'|$, which reflects the spatial homogeneity and isotropy of de Sitter space-time in this coordinate system.

It proves convenient to introduce conformally rescaled quantities, such as the rescaled field $\phi(x)=a^{d-1\over2}(\eta)\varphi(x)$ and its correlators. One has, for instance, 
\beq
\label{eq:rep1}
 F(x,x')=\left[a(\eta)a(\eta')\right]^{-{d-1\over2}}F_c(\eta,\eta',|\bX-\bX'|),
\eeq
where $F_c(\eta,\eta',|\bX-\bX'|)={1\over2}\langle \{\phi(x),\phi(x')\}\rangle$. Introducing spatial comoving momentum variables, one defines
\beq
 \tilde F_c(\eta,\eta',K)=\int d^d S\, e^{-i\bK\cdot\bS}F_c(\eta,\eta',|\bS|).
\eeq
Spatial homogeneity implies the conservation of the comoving momentum.

Specializing to de Sitter, an extra symmetry implies that two-point correlators follow the scaling law \cite{Parentani:2012tx,Adamek:2013vw} 
\beq
\label{eq:prep}
 \tilde F_c(\eta,\eta',K)=\frac{1}{K}\hat F(p,p'),
\eeq
where $p=-K\eta$ and $p'-K\eta'$ are the physical momenta at times $\eta$ and $\eta'$ respectively. \Eqn{eq:prep} provides the $p$-representation of the two-point correlator.

\begin{figure}[t!]  
\epsfig{file=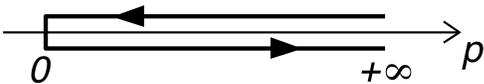,width=6.5cm}
 \caption{\label{fig:ppath} 
The closed path $\hat\C=\hat\C^+\cup\hat\C^-$ in the momentum variable $p=-K\eta$. The upper branch $\hat\C^+$ goes from $+\infty$ to $0^+$ and the lower branch $\hat\C^-$ goes back from $0^+$ to $+\infty$.}
\end{figure}

As mentioned previously \Eqn{eq:prep} applies to the spectral function $\rho$ as well. In fact, as in \eqn{eq:ttt}, one can combine the statistical and spectral two-point functions in a single propagator defined on a closed contour $\hat\C$ in physical momentum \cite{Parentani:2012tx}:
\beq
\label{eq:G}
 \hat G(p,p')=\hat F(p,p')-\frac{i}{2}{\rm sign}_{\hat\C}(p-p')\hat\rho(p,p').
\eeq
The sign function is to be understood on the contour depicted in Fig.~\ref{fig:ppath}. We also note the symmetry properties $F(p,p')=F(p',p)$ and $\rho(p,p')=-\rho(p',p)$.

\section{$O(N)$ scalar field theory in the large-$N$ limit}
\label{sec:ON}

We consider an $O(N)$ scalar theory, defined by the classical action
\beq
\label{eq:theory}
 {\cal S}[\varphi]=\int_x\left\{{1\over2}\varphi_a\left(\square-m_{\rm dS}^2\right)\varphi_a-\frac{\lambda}{4!N}(\varphi_a\varphi_a)^2\right\},
\eeq
where $a=1,\ldots,N$ and a summation over repeated indices is understood. Here, 
\beq
 \square\equiv\frac{1}{\sqrt{-g(x)}}\partial_\mu\sqrt{-g(x)}g^{\mu\nu}\partial_\nu
\eeq
is the covariant Laplace operator and 
\beq
\label{eq:mds}
 m_{\rm dS}^2=m^2+\xi R=m^2+d(d+1)\xi
\eeq
is the effective square mass with $m$ the tree-level mass and $\xi$ the coupling to the Ricci scalar $R=d(d+1)$. Finally, 
\beq
\label{eq:measure}
 \int_x\equiv\int d^Dx\sqrt{-g(x)}=\int_\C dx^0\int d^dx\sqrt{-g(x)},
\eeq 
where the $x^0$ integral runs along the contour $\C$.

\subsection{The large-$N$ limit}

\begin{figure}[t!]  
\epsfig{file=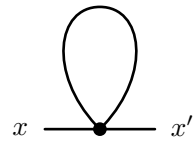,width=2.5cm}
 \caption{\label{fig:Sigma} 
The self-energy $\Sigma(x,x')$ in the limit $N\to\infty$. The internal line in the diagram is given by the full propagator $G(x,x)$, hence the nonperturbative nature of the large-$N$ limit. This describes self-consistent mass generation.}
\end{figure}
 
The $1/N$ expansion provides a controlled expansion scheme which gives access to nonperturbative aspects of the theory. It can be formulated in cosmological spaces \cite{Ramsey:1997qc,Parentani:2012tx} using standard functional techniques, provided one employs appropriate covariant quantities such as the functional derivative,
\beq
 {\delta_c\over\delta\varphi(x)}\equiv {1\over \sqrt{-g(x)}}{\delta\over\delta\varphi(x)},
\eeq 
the integration measure \eqn{eq:measure} and the corresponding covariant Dirac distribution on the contour,
\beq
 \delta^{(D)}(x,x')=\frac{\delta_\C(x^0-x^{\prime0})\delta^{(d)}(\bx-\bx')}{\sqrt{-g(x)}},
\eeq
defined such that $\int_z\delta^{(D)}(x,z)f(z)=f(x)$ for any function $f$ on the contour. 
Let us recall the basic equations describing the limit $N\to \infty$, see \cite{Aarts:2002dj,Parentani:2012tx}. It has been shown in \cite{Serreau:2011fu} that the theory does not admit a de Sitter invariant state with spontaneous symmetry breaking. We thus consider the symmetric phase, where $\langle\varphi_a\rangle=0$, $G_{ab}=\delta_{ab}G$ and  $\Sigma_{ab}=\delta_{ab}\Sigma$. The covariant inverse propagator $G^{-1}$ is defined as
\beq
 \int_z G^{-1}(x,z)G(z,x')=\delta^{(D)}(x,x').
\eeq
It can be written in terms of the covariant self-energy as
\beq
 G^{-1}(x,x')=G_0^{-1}(x,x')-\Sigma(x,x'),
\eeq
where 
\beq
\label{eq:freeprop}
 iG_0^{-1}(x,x')=(\square_x-m_{\rm dS}^2)\delta^{(D)}(x,x').
\eeq
In the large-$N$ limit, the self-energy is given by the tadpole diagram of Fig. \ref{fig:Sigma}, where the line represents the propagator $G$ itself. It thus corresponds to a local, masslike contribution:
\beq
 \Sigma(x,x')=-i\sigma\delta^{(D)}(x,x'),
\eeq
where
\beq
\label{eq:masscorrec}
 \sigma=\frac{\lambda}{6}G(x,x).
\eeq
De Sitter symmetry guarantees that the latter is a constant. The propagator $G$ thus satisfies the equation
\beq
\label{eq:propeq}
 \left(\square_x-M^2\right)G(x,x')=i\delta^{(D)}(x,x'),
\eeq
with a self-consistent square mass
\beq
\label{eq:mass}
 M^2=m_{\rm dS}^2+\sigma.
\eeq
 
\begin{figure}[t!]  
\epsfig{file=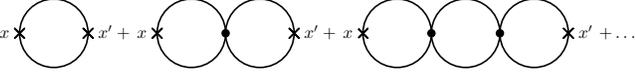,width=8.5cm}
 \caption{\label{fig:Ifunc} 
The infinite series of bubble diagrams contributing to the function $I(x,x')$. The black dots denote interaction vertices whereas the crosses are the end points of the function. The elementary one-loop bubble is given by the function $\Pi(x,x')\propto G^2(x,x')$. Each addition of a new bubble leads to large IR logarithms in the momentum representation, the infinite series of which resums to a modified power law.}
\end{figure}
The four-point vertex function reads, writing  $x_i\equiv x_1,\ldots,x_4$ for brief,
\bea
\label{eq:gamma4}
 \Gamma^{(4)}_{abcd}(x_i)\!\!&=&\!\!\delta_{ab}\delta_{cd}\,\delta^{(D)}(x_1,x_2)\delta^{(D)}(x_3,x_4) iD(x_1,x_3)\nn
 &+&{\rm perm.}\,,
\eea
where ``perm.'' stands for the two cyclic permutations needed to make $\Gamma^{(4)}_{abcd}(x_i)$ symmetric. Here, the function
\beq
\label{eq:D}
 iD(x,x')=-{\lambda\over3N}\left[\delta^{(D)}(x,x')+iI(x,x')\right]
\eeq
is the two-point correlator of the composite field $\chi=\lambda\varphi_a\varphi_a/6N$: $D(x,x')=\langle T_\C\chi(x)\chi(x')\rangle$.
The local contribution gives rise to the classical vertex in \eqn{eq:gamma4} and the nonlocal one actually corresponds to the infinite sum of bubble diagrams represented in Fig. \ref{fig:Ifunc}. This infinite sum can be cast into the following integral equation: 
\beq
\label{eq:I}
 {I}(x,x')=\Pi(x,x')+i\int_z\Pi(x,z){I}(z,x'),
\eeq
with 
\beq
\label{eq:Pi}
 \Pi(x,x')=-{\lambda\over6}G^2(x,x')
\eeq
the elementary one-loop bubble diagram. The four-point vertex is thus completely expressed in terms of a nontrivial two-point function.

 \subsection{$p$-representation}

The $p$-representation plays an important role in the analytical solution presented below. We briefly review the relevant material for our present purposes and refer the reader to Ref. \cite{Parentani:2012tx} for more details.

\Eqn{eq:propeq} is equivalent to the free-field-like equation for the statistical function $\hat F$ defined in Eqs. \eqn{eq:rep1}-\eqn{eq:prep}
\beq
\label{eq:eom}
\left[\partial_p^2+1-\frac{\nu^2-{1\over4}}{p^2}\right] \hat F(p,p')=0,
\eeq
where
\beq
\label{eq:nu}
 \nu=\sqrt{{d^2\over4}-M^2}.
\eeq
The spectral function $\hat\rho(p,p')$ satisfies the same equation. The ``initial'' data is to be specified at $p=p'\to\infty$. Commutation relations imply $\hat\rho(p,p')|_{p=p}=\partial_p\partial_{p'}\hat\rho(p,p')|_{p=p'}=0$ and $\partial_p\hat\rho(p,p')|_{p=p'}=-1$ for the spectral function. The statistical function contains the information about the (quantum) state of the system. The only choice compatible with de Sitter symmetry and with the requirement of renormalizability is the Bunch-Davies vacuum \cite{Bunch:1978yq}. This corresponds to the choice
\bea
 \left.\hat F(p,p')\right|_{p=p'\to\infty}&=&{1\over2},\nn
 \left.\partial_p\hat F(p,p')\right|_{p=p'\to\infty}&=&0,\\
 \left.\partial_p\partial_{p'}\hat F(p,p')\right|_{p=p'\to\infty}&=&{1\over2}.\nonumber
\eea

With this choice, the solutions of \Eqn{eq:eom} for $\hat F$ and $\hat \rho$ read
\bea
 \hat F(p,p')&=&{\pi\over4}\sqrt{pp'}\,{\rm Re}\left\{H_\nu(p)H^*_\nu(p')\right\}\\
 \hat\rho(p,p')&=&-{\pi\over2}\sqrt{pp'}\,{\rm Im}\left\{H_\nu(p)H^*_\nu(p')\right\},
\eea
where $H_\nu(x)$ is the Hankel function of the first kind \cite{Gradshteyn}.
Finally, the self-energy \eqn{eq:masscorrec} reads 
\beq
\label{eq:sigma}
 \sigma=\frac{\lambda}{6}\int_\bq\frac{\hat F(q,q)}{q}=\frac{\lambda\pi}{24}\int_\bq\left|H_\nu(q)\right|^2,
\eeq
where $\int_\bq=\int\frac{d^dq}{(2\pi)^d}$. The self-consistent mass $M$ is the solution of the gap equation \eqn{eq:mass} with
\eqn{eq:sigma} and \eqn{eq:nu}. For instance, in the case of vanishing or negative tree-level (classical) mass $m_{\rm dS}^2\le0$, where perturbation theory is ill-defined, \Eqn{eq:mass} describes the dynamical generation of a strictly positive square mass which cures the IR divergences of perturbation theory.

Let us now consider the four-point vertex function \eqn{eq:gamma4}. Introducing the appropriate conformal factors
\beq
 \Gamma^{(4)}_{abcd}(x_i)=\left[a(\eta_1)\cdots a(\eta_4)\right]^{-{d+3\over2}}\Gamma^{(4)}_{c,abcd}(x_i)
\eeq
and going to spatial comoving momentum space, we have, extracting a factor $(2\pi)^d\delta^{(d)}\left(\sum_{i=1}^4\bK_i\right)$,
\bea
 &&\hspace{-0.4cm}\tilde\Gamma^{(4)}_{c,abcd}(\eta_i,\bK_i)=\left[a(\eta_1)\cdots a(\eta_4)\right]^{3-d\over4}\nn
 &&\hspace{-0.4cm}\times\Big\{\delta_{ab}\delta_{cd}\,\delta_\C(\eta_1\!-\!\eta_2)\delta_\C(\eta_3\!-\!\eta_4)i\tilde D_c(\eta_1,\eta_3,K_{12})\!+\!{\rm perm.}\!\Big\},\nn
 \label{eq:fourpointconf}
\eea
where $K_{ij}=|\bK_i+\bK_j|$. Here,  $\tilde D_c$ is the comoving momentum representation of the function $D$, \Eqn{eq:D}, defined as
\beq
\label{eq:Drep1}
 D(x,x')=\left[a(\eta)a(\eta')\right]^{-{d+1\over2}}D_c(\eta,\eta',|\bX-\bX'|)
\eeq
and
\beq
\label{eq:Drep2}
  \tilde D_c(\eta,\eta',K)=\int d^d S\, e^{-i\bK\cdot\bS}D_c(\eta,\eta',|\bS|).
\eeq
We define the comoving momentum representations $\tilde I_c$ and $\tilde\Pi_c$ of the functions $I$ and $\Pi$, Eqs. \eqn{eq:I} and \eqn{eq:Pi}, in the same way as for $\tilde D_c$. Equation \eqn{eq:D} then reads
\beq
 i\tilde D_c(\eta,\eta',K)=-{\lambda\over3N}\left[\delta_\C(\eta-\eta')+i\tilde I_c(\eta,\eta',K)\right].
\eeq
As already emphasized, the function $\tilde I_c$ encodes the nontrivial (loop) contribution to the four-point vertex in the large-$N$ limit. It corresponds to the infinite sum of bubble diagrams depicted in Fig.~\ref{fig:Ifunc} with the elementary one-loop bubble $\tilde\Pi_c$.
The functions $\tilde D_c$ and $\tilde \Pi_c$ and $\tilde I_c$ admit the following $p$-representation:
\beq
 \tilde D_c(\eta,\eta',K)=K\hat D(p,p')
\eeq
and similarly for $\tilde \Pi_c$ and $\tilde I_c$, where $p=-K\eta$ and $p'=-K\eta'$. Introducing the Dirac distribution on the momentum contour\footnote{The latter is defined such that $\int_{\hat{\C}}dp' \delta_{\hat{\C}}(p-p')\hat f(p')=\hat f(p)$ for any function $\hat f$ on the contour $\hat{\C}$. The minus sign in \eqn{eq:delta} copes for the ``wrong'' orientation of the momentum contour. Note, in particular,  that $\int_{\hat{\C}} dp=-K\int_\C d\eta$ \cite{Parentani:2012tx}.} $\hat\C$ 
\beq
\label{eq:delta}
 \delta_\C(\eta-\eta')=- K \delta_{\hat\C}(p-p'),
\eeq
one has
\beq
\label{eq:Dprep}
 i\hat D(p,p')=-{\lambda\over3N}\left[-\delta_{\hat\C}(p-p')+i\hat I(p,p')\right].
\eeq
The one-loop bubble $\hat\Pi$ is given by
\beq
\label{eq:Np1}
 \hat\Pi(p,p')=-\frac{\lambda}{6}\,(pp')^{d-3\over2}\!\!\int_\bq\frac{\hat G\left(qp,qp'\right)}{q}\frac{\hat G\left(rp,rp'\right)}{r},
\eeq
where $r=|\be+\bq|$, with $\be$ an arbitrary unit vector, and the function $\hat I$ solves the following integral equation on the contour $\hat\C$:
\beq
\label{eq:Np2}
 \hat I(p,p')=\hat\Pi(p,p')-i\int_{\hat\C} ds \,\hat\Pi(p,s )\hat I(s ,p').
\eeq

Finally, we write the above equations explicitly in terms of their statistical and spectral components on the contour, defined as\footnote{This decomposition is modified by renormalization. However, the final equations to be discussed here are unchanged, see Appendix \ref{appsec:renorm}.}
\beq
\label{eq:Pidec}
 \hat\Pi(p,p')=\hat \Pi_F(p,p')-\frac{i}{2}{\rm sign}_{\hat\C}(p-p')\hat\Pi_\rho(p,p'),
\eeq
with $\Pi_F(p,p')=\Pi_F(p',p)$ and $\Pi_\rho(p,p')=-\Pi_\rho(p',p)$ and similarly for $\hat I$. One has
\bea
\label{eq:PiF}
 \hat\Pi_F(p,p')&=&-\frac{\lambda}{6}\,(pp')^{d-3\over2}\!\!\int_\bq\frac{1}{qr}\Big\{\hat F\!\left(qp,qp'\right)\!\hat F\!\left(rp,rp'\right)\nn
 &&-{1\over4}\hat \rho\!\left(qp,qp'\right)\!\hat \rho\!\left(rp,rp'\right)\Big\}
\eea
and
\beq
\label{eq:Pirho}
 \hat\Pi_\rho(p,p')=-\frac{\lambda}{3}\,(pp')^{d-3\over2}\!\!\int_\bq\frac{\hat F\!\left(rp,rp'\right)\hat \rho\!\left(rp,rp'\right)}{qr}.
\eeq
Writing explicitly the integrals on the contour $\hat\C$ in \eqn{eq:Np2}, one obtains the following equations
\beq
\label{eq:Irho}
 \hat I_\rho(p,p')=\hat\Pi_\rho(p,p')+\int^{p'}_{p} ds \,\hat\Pi_\rho(p,s )\hat I_\rho(s ,p')
\eeq
and
\beq
\label{eq:IF}
 \hat I_F(p,p')=\hat\Pi_H(p,p')+\!\int_{p}^\infty \!\!ds \,\hat\Pi_\rho(p,s )\hat I_F(s ,p'),
\eeq
where we introduced the auxiliary function 
\beq
\label{eq:PiH}
 \hat \Pi_H(p,p')=\hat\Pi_F(p,p')-\!\int_{p'}^\infty \!\!ds \,\hat\Pi_F(p,s )\hat I_\rho(s ,p').
\eeq
For later use, we note that, using \eqn{eq:Irho}, \Eqn{eq:IF} can be solved in terms of the function $\hat I_\rho$ as, see Appendix \ref{appsec:proof},
\beq
\label{eq:IF2}
 \hat I_F(p,p')=\hat\Pi_H(p,p')+\!\int_{p}^\infty \!\!ds \,\hat I_\rho(p,s )\hat \Pi_H(s ,p').
\eeq

In the next two sections, we solve the above equations in the deep IR limit, that is for $p,p'\ll1$ in the case of light fields, $M^2\ll1$. We adopt the following general strategy: we first assume that the IR physics is dominated by IR modes and we cut all momentum integrals (either in momentum loops or in ``time'' integrals) at a momentum scale $\mu\lesssim1$. We shall see that we can find a closed solution of the equations in that case and we verify {\it a posteriori} that the solution weakly depends on the scale $\mu$. Of course this is not a proof of our assumption, merely a consistency check. We defer to the Appendices a more complete analysis taking into account the role of high momentum modes. There we show that our results only get slightly modified by a constant renormalization factor. Similarly, we neglect renormalization aspects in the main course of the text and defer their discussion to Appendix \ref{appsec:renorm}.

\section{Two-point function: mass generation}
\label{sec:2pt}

The solution of the gap equation \eqn{eq:mass} has been discussed in \cite{Riotto:2008mv,Garbrecht:2011gu,Serreau:2011fu}. Here, we briefly review the main results and introduce some necessary material for the next section, where we compute the four-point function.

In the case of light fields, with $M<d/2$, the index $\nu$ is real and $0<\nu<d/2$. It is customary to introduce the parameter $\varepsilon=d/2-\nu$. For small mass, one has $\epsilon\approx M^2/d$.
Following the strategy described above, we only need the statistical and spectral correlators $\hat F$ and $\hat\rho$ in the IR regime, i.e., for $p,p'\ll1$. The latter are given by, see Appendix \ref{appsec:tpt},
\beq
\label{eq:FIR}
 \hat F_{\rm IR}(p,p')=\frac{F_\nu}{(pp')^{\nu-{1\over2}}},
\eeq
where $F_\nu=[2^\nu\Gamma(\nu)]^2/4\pi$ and
\beq
\label{eq:rhoIR}
 \hat \rho_{\rm IR}(p,p')=-\frac{\sqrt{pp'}}{2\nu}\left[\left({p\over p'}\right)^{\!\!\nu}\!-\left({p'\over p}\right)^{\!\!\nu}\right].
\eeq
It proves useful to rewrite the latter expression as
\beq
\label{eq:rhoIR2}
  \hat \rho_{\rm IR}(p,p')=-\sqrt{pp'}\,{\cal P}_\nu\left(\ln{p\over p'}\right),
\eeq
with
\beq
\label{eq:P}
 {\cal P}_\nu(x)=\frac{\sinh(\nu x)}{\nu}.
\eeq

Restricting the momentum integration in \eqn{eq:sigma} to the IR part $|\bq|<\mu$, one gets
\beq
 \sigma=\frac{\lambda}{6}\int_{|\bq|<\mu}\!\!\!\frac{\hat F_{\rm IR}(q,q)}{q}\approx\frac{\lambda F_\nu}{12\varepsilon}\frac{\Omega_d}{(2\pi)^d},
\eeq
up to corrections of relative order $\varepsilon\ln\mu\ll1$. Here $\Omega_d=2\pi^{d/2}/\Gamma(d/2)$. Using $\epsilon\approx M^2/d$ in the small mass case, it is easy to solve the gap equation \eqn{eq:mass}. For instance, in the case of vanishing tree-level mass, $m_{\rm dS}^2=0$, one gets
\beq
\label{eq:masssss}
 M^2=\sqrt{\frac{\lambda F_\nu}{12}\frac{d\Omega_d}{(2\pi)^d}}
\eeq
so that
\beq
\label{eq:epsilon}
 \varepsilon={\sigma\over d}={\alpha(d/2)\over12\pi}\sqrt\lambda
\eeq
where $\alpha(x)=\sqrt{{3\Gamma(x)/x\pi^{x-1}}}$ is chosen here so that $\alpha(3/2)=1$ for $d=3$. The self-consistent square mass \eqn{eq:masssss} is nonanalytic in the coupling, reflecting its nonperturbative IR character.

\section{Four-point function: resumming infrared logarithms}
\label{sec:four}

The first step in the calculation of the four-point vertex is the evaluation of the bubble $\hat\Pi$, Eqs. \eqn{eq:PiF} and \eqn{eq:Pirho}. Restricting the loop integral to the contribution from IR modes, i.e. such that $qp,qp'\lesssim\mu$, which also implies $rp,rp'\lesssim\mu$, one can replace the propagators $\hat F$ and $\hat\rho$ in Eqs. \eqn{eq:PiF} and \eqn{eq:Pirho} by their IR expressions \eqn{eq:FIR}--\eqn{eq:P}. The loop integrals are performed in Appendix \ref{appsec:piIR}. We obtain
\beq
\label{eq:PiFIR}
 \hat\Pi_F^{\rm IR}(p,p')=\frac{\pi_F}{(pp')^{\kappa+{1\over2}}},
\eeq
with $\kappa={d\over2}-2\varepsilon$ (notice that $\kappa=\nu-\varepsilon<\nu$) and
\beq
\label{eq:PirhoIR}
 \hat\Pi_\rho^{\rm IR}(p,p')=\frac{\pi_\rho}{\sqrt{pp'}}\,{\cal P}_\nu^\varepsilon\!\left(\ln{p\over p'}\right),
\eeq
where we defined the odd function
\beq
 {\cal P}_\nu^\varepsilon(x)=e^{-\varepsilon |x|}\,{\cal P}_\nu(x),
\eeq
with ${\cal P}_\nu(x)$ given in \eqn{eq:P}, and where 
\beq
\label{eq:relation}
 \pi_\rho=\frac{\lambda F_\nu}{6\varepsilon}\frac{\Omega_d}{(2\pi)^d}=2\sigma\quad{\rm and}\quad\pi_F=-\pi_\rho F_\nu.
\eeq

Let us first consider \Eqn{eq:Irho} for $\hat I_\rho$. We seek a solution of the form
\beq
\label{eq:Iansatz}
 \hat I_\rho^{\rm IR}(p,p')=\frac{\pi_\rho}{\sqrt{pp'}}\,{\cal I}_\varepsilon\!\left(\ln{p\over p'}\right),
\eeq
where
\beq
\label{eq:calIeps}
 {\cal I}_\varepsilon(x)=e^{-\varepsilon |x|}\,{\cal I}(x).
\eeq
with ${\cal I}(x)$ an unknown odd function, as implied by the antisymmetry of $\hat I_\rho(p,p')$ in the exchange $p\leftrightarrow p'$. The $\exp(-\varepsilon |x|)$ terms cancel out in \Eqn{eq:Irho} and we are left with the following integral equation:
\beq
\label{eq:calI}
 {\cal I}(x)={\cal P}_\nu(x)-\pi_\rho\int_0^x\!\!dy\,{\cal P}_\nu(x-y)\,{\cal I}(y).
\eeq

As a first attempt to solve this equation, one may try an expansion in the parameter $\pi_\rho$ which counts the number of bubbles. This is akin to perturbation theory. At first order, one has
\bea
 {\cal I}(x)&=&{\cal P}_\nu(x)-\pi_\rho\int_0^x\!\!dy\,{\cal P}_\nu(x-y)\,{\cal P}_\nu(y)+{\cal O}\left(\pi_\rho^2\right)\nn
 &=&{\cal P}_\nu(x)\left(1+{\pi_\rho\over2\nu}\left[1-{x\over\tanh(\nu x)}\right]+{\cal O}\left(\pi_\rho^2\right)\right).\nn
\eea
For $|x|\gg1$, that is for large logarithms or, equivalently, large cosmological time separations, $|\ln(p/p')|=|t-t'|\gg1$, we get
\beq
 {\cal I}(x)\approx{\cal P}_\nu(x)\left(1-{\pi_\rho\over2\nu}|x|+{\cal O}\left(\pi_\rho^2\right)\right).
\eeq
Thus, we see that the true expansion parameter is $\pi_\rho|x|=\pi_\rho|\ln(p/p')|$ and that the perturbative solution breaks down for $|\ln(p/p')|=|t-t'|\gtrsim1/\pi_\rho$. The argument extends to higher orders and one can check that, in that case, one has to resum the infinite series of bubble diagrams.

Fortunately, the exact solution of \Eqn{eq:calI} is easily obtained. It reads
\beq
\label{eq:calIsol}
 {\cal I}(x)=\frac{\sinh\bar\nu x}{\bar\nu}={\cal P}_{\bar\nu}(x),
\eeq
where (note that $\bar\nu<\nu$)
\beq
\label{eq:nubar}
 \bar\nu=\sqrt{\nu^2-\pi_\rho}\,.
\eeq
We thus have
\beq
\label{eq:Irhosol}
 \hat I_\rho^{\rm IR}(p,p')=\frac{\pi_\rho}{\sqrt{pp'}}\,{\cal P}_{\bar\nu}^{\varepsilon}\!\left(\ln{p\over p'}\right).
\eeq
We see that resumming the infinite series of bubble diagrams---each producing higher powers of $\pi_\rho|\ln p/p'|$---through the integral equation \eqn{eq:Irho} changes the exponent $\nu$ characterizing the IR power law behavior of the single bubble function $\hat\Pi_\rho$, see \Eqn{eq:PirhoIR}, into the new exponent \eqn{eq:nubar}. 

Let us now consider the case of $\hat I_F$, \Eqn{eq:IF}, where, following the general strategy described previously, we replace the upper bound of momentum integrals in \eqn{eq:IF}--\eqn{eq:IF2} by $\mu\lesssim1$. The contribution of higher momenta is analyzed in Appendix~\ref{appsec:roleofUV}. We first note that the $F$ component of the symmetric bubble function $\hat\Pi$ can be written as the following factorized form
\beq
\label{eq:factorization}
 \hat\Pi_F^{\rm IR}(p,p')=\frac{\pi_F}{\sqrt{pp'}}\,{\cal A}\left(\ln {p\over\mu}\right){\cal A}\left(\ln {p'\over\mu}\right),
\eeq
where we artificially introduced a dependence on $\mu$ for later convenience. Here,
\beq
\label{eq:Adef}
 {\cal A}(x)=\frac{e^{-\kappa x}}{\mu^\kappa}.
\eeq
The factorization property \eqn{eq:factorization} together with the fact that \Eqn{eq:PiH} for $\hat\Pi_H^{\rm IR}(p,p')$ only involves the function $\hat\Pi_F^{\rm IR}(p,s)$ with fixed first argument imply that the auxiliary function takes the asymmetric factorized form,
\beq
 \hat\Pi_H^{\rm IR}(p,p')=\frac{\pi_F}{\sqrt{pp'}}\,{\cal A}\left(\ln {p\over\mu}\right)\bar {\cal A}\left(\ln {p'\over\mu}\right),
\eeq
with the function ${\cal A}$ given in \eqn{eq:Adef} above and the function $\bar{\cal A}$ is to be determined.
Using \eqn{eq:PiH} and \eqn{eq:Iansatz}, we find (here $x=\ln p/\mu<0$)
\beq
\label{eq:abareq}
 \bar {\cal A}(x)={\cal A}(x)+\pi_\rho\int_x^0\!\!dy \,{\cal I}_{\varepsilon}(x-y)\,{\cal A}(y),
\eeq
where we used $I_{\varepsilon}(x)=-I_{\varepsilon}(-x)$. We get
\beq
\label{eq:abarsol}
 \bar {\cal A}(x)={\pi_\rho\over2\bar\nu}{\cal A}(x)\left(\frac{e^{(\nu-\bar\nu)x}}{\nu-\bar\nu}-\frac{e^{(\nu+\bar\nu)x}}{\nu+\bar\nu}\right).
\eeq

Finally, using the solution \eqn{eq:IF2} of \Eqn{eq:IF}, where the function $\hat\Pi_H^{\rm IR}(s,p')$ always appeasr with the same second argument as $\hat I_F^{\rm IR}(p,p')$, we conclude that the latter also factorizes:
\beq
\label{eq:IFsol}
 \hat I_F^{\rm IR}(p,p')=\frac{\pi_F}{\sqrt{pp'}}\,\bar {\cal A}\left(\ln {p\over\mu}\right)\bar {\cal A}\left(\ln {p'\over\mu}\right),
\eeq
with the function $\bar{\cal A}$ given in \Eqn{eq:abarsol} above. The factorization property \eqn{eq:IFsol} of $ \hat I_F^{\rm IR}$ follows directly from the factorization of $\hat\Pi_F^{\rm IR}$, \Eqn{eq:factorization}. As described in Appendix \ref{appsec:piIR}, the latter follows from the corresponding factorization property \eqn{eq:FIR} of the statistical correlator $\hat F_{\rm IR}$ and is thus reminiscent of the fact that IR de Sitter fluctuations can be described as a classical stochastic ensemble of growing modes\footnote{The fact that the connected two-point function factorizes is also to be put in parallel with the phenomenon of undulation in (analog) black hole physics \cite{Coutant:2012zh}. In the present case, the undulation occurs in physical momentum space.} \cite{Guth:1985ya,Polarski:1995jg,Campo:2003pa}.

Let us discuss in more detail the above solution, \Eqn{eq:IFsol}. Recalling that $\bar\nu<\nu$, one finds, for $-x\gg1$, 
\beq\label{eq:abar}
 \bar {\cal A}(x)\approx{\pi_\rho\mu^{\bar\nu-\nu}\over2\bar\nu(\nu-\bar\nu)}\frac{e^{-\bar\kappa x}}{\mu^{\bar\kappa}}\approx\frac{e^{-\bar\kappa x}}{\mu^{\bar\kappa}},
\eeq
where we neglected terms of relative order ${\cal O}(\pi_\rho)$ in the second equality. Here, we introduced the new exponent
\beq
\label{eq:kbar}
 \bar\kappa=\bar\nu-\varepsilon.
\eeq
Again, we find that the infinite summation of bubble diagrams changes the IR power law of the single bubble
\beq
  {\cal A}\left(\ln {p\over\mu}\right)={1\over p^{\kappa}}\quad\longrightarrow\quad\bar {\cal A}\left(\ln {p\over\mu}\right)\approx{1\over p^{\bar\kappa}}
\eeq
and $\bar\kappa-\kappa$ can be regarded as an anomalous dimension. We thus have
\beq
\label{eq:IFresult}
 \hat I_F^{\rm IR}(p,p')\approx\frac{\pi_F}{(pp')^{\bar\kappa+{1\over2}}}.
\eeq
We emphasize that the dependence of the above results on the scale $\mu\lesssim 1$ is quite mild $\propto\mu^{\bar\nu-\nu}\approx1-{\pi_\rho\over2\nu}\ln\mu$, in the case $\pi_\rho\ll1$. This suggests that the neglect of high momentum modes in the present analysis is a consistent approximation. 

As in the case of $\hat I_\rho$, it is instructive to compare the exact solution \eqn{eq:abar} to the perturbative expansion in powers of $\pi_\rho$. Using \eqn{eq:calI}, \Eqn{eq:abareq} reads, at lowest nontrivial order,
\bea
 \bar {\cal A}(x)&=&{\cal A}(x)+\pi_\rho\int^0_x\!\!dy \,{\cal P}_{\varepsilon}(x-y)\, {\cal A}(y)+{\cal O}\left(\pi_\rho^2\right)\nn
 &=&{\cal A}(x)\left(1+{\pi_\rho\over2\nu}\left[x+{1\over2\nu}-\frac{e^{2\nu x}}{2\nu}\right]+{\cal O}\left(\pi_\rho^2\right)\right).\nn
\eea
For IR momenta, $-x=\ln(\mu/p)\gg1$, we get
\beq
 \bar {\cal A}(x)\approx {\cal A}(x)\left(1+{\pi_\rho\over2\nu}x+{\cal O}\left(\pi_\rho^2\right)\right).
\eeq
We see that the true expansion parameter is, in fact, $\pi_\rho\ln\mu/p$ and that, for momenta $\ln(\mu/p)\gtrsim1/\pi_\rho$, one needs to resum the IR logarithms generated by the series of bubble diagrams. This results in the modification of the IR exponent $\kappa\to\bar\kappa$.

\section{Discussion}
\label{sec:discussion}

In the previous sections, we have mainly considered the case of a small, or vanishing, tree-level square mass $m_{\rm dS}^2\to0$. In this  case, perturbation theory suffers from strong IR divergences. As previously emphasized, one has, in that case, $\varepsilon\propto\sqrt\lambda$ and thus $\pi_\rho\propto\sqrt\lambda$. The actual expansion parameter is thus $\sqrt{\lambda}$. The mass generation phenomenon discussed in Sec. \ref{sec:2pt} can be viewed as a sort of anomalous dimension in the IR \cite{Boyanovsky:2012qs}. Indeed the IR power law behavior of the free propagator of a massless field, characterized by the exponent $d/2$, is changed to an exponent $\nu\approx d/2-\varepsilon$ which regulates the IR divergences of the perturbative series.

Similarly, the resummation of nonlocal IR logarithms in the two-point function of the composite operator $\varphi^2$ or, equivalently, in the four-point vertex function of the field $\varphi$ leads to a modified IR behavior: the power law behaviors of the one-loop bubble contributions $\hat \Pi_\rho$ and $\hat \Pi_F$, characterized by the exponents $\nu\approx d/2-\varepsilon$ and $\kappa\approx d/2-2\varepsilon$, are changed to $\bar\nu\approx d/2-3\varepsilon$ and $\bar\kappa\approx d/2-4\varepsilon$ respectively. 

We point out that, inversely, the dynamically generated IR power law behavior of the function $\hat D$, for large (super-Hubble) momentum/time separations $|\ln p/p'|=|t-t'|\gtrsim 1$, can be viewed as the generation of an effective mass for the composite field $\varphi^2$.  Using, see \eqn{eq:kbar},
\beq
 e^{-\varepsilon|x|}\sinh \bar\nu x\approx \sinh \bar\kappa x, \quad{\rm for}\quad |x|\gtrsim1,
\eeq
as well as the relation \eqn{eq:relation} between $\pi_F$ and $\pi_\rho$, one has, for $|\ln p/p'|\gtrsim 1$,
\bea
 pp'\hat D_F^{\rm IR}(p,p')&\approx&\frac{\lambda\pi_\rho}{3N}\sqrt{pp'}\frac{F_{\bar\kappa}}{(pp')^{\bar\kappa}}\\
 pp'\hat D_\rho^{\rm IR}(p,p')&\approx&-\frac{\lambda\pi_\rho}{3N}\sqrt{pp'}{\cal P}_{\bar \kappa}\!\left(\ln {p\over p'}\right).
\eea
We have assumed $\varepsilon\ll1$ and $\pi_\rho\ll1$. In this case, one can identify $\bar\kappa\approx\bar\nu\approx\nu$ in the constant factors on the right-hand side. Using \Eqn{eq:rhoIR2}, we see that the function $pp'\hat D^{\rm IR}(p,p')$ behaves as the two-point function of a free massive field with square mass $\bar M^2=d^2/4-\bar\kappa^2$. For vanishing tree-level mass, we get $\bar M\approx 2M$.

There is another interesting situation to be mentioned, namely the case of a negative tree-level square mass $m_{\rm dS}^2<0$. This is the case where the classical potential shows spontaneous symmetry breaking. In that case, perturbation theory is again ill-defined because of the presence of massless Goldstone modes. It has been shown in \cite{Serreau:2011fu} that, as in the previous case, nonperturbative effects of self-interactions cure the theory by generating a positive mass square restoring the symmetry. The self-consistent mass is approximately given by 
\beq
 M^2={m_{\rm dS}^2\over2}+\sqrt{{\left(m_{\rm dS}^2\right)^2\over4}+\frac{\lambda F_\nu}{12}\frac{d\Omega_d}{(2\pi)^d}}\approx\frac{\lambda F_\nu}{12|m_{\rm dS}^2|}\frac{d\,\Omega_d}{(2\pi)^d},
\eeq
from which it follows that 
\beq
 \varepsilon=\frac{\lambda F_\nu}{12|m_{\rm dS}^2|}\frac{d\,\Omega_d}{(2\pi)^d}
\eeq
and thus 
\beq
 \pi_\rho=2|m_{\rm dS}^2|.
\eeq
This is equal to the radial curvature of the tree-level potential at its minimum.
We see that, in that case, $\epsilon\propto\lambda$ is parametrically smaller than in the case of vanishing tree-level mass, leading to stronger IR enhancement and to a truly nonperturbative value of the parameter $\pi_\rho$ governing the bubble resummation. At weak coupling, we get $\bar M\approx \sqrt \pi_\rho\gg M$. The composite field $\varphi^2$ gets a nonperturbatively large mass.

\section{Conclusions}

Resumming perturbative IR divergences of QFT on de Sitter space is a timely issue of modern cosmology. The present study provides an exact solution of a nonperturbative limit where large IR logarithms can be explicitly resummed. This gives rise to modified power laws for the correlators in the IR, akin to anomalous dimensions in the context of critical phenomena. 

A detailed analysis of the role of UV modes reveals that they only affect the IR results by a multiplicative constant. This is a form of effective decoupling between UV and IR physics in de Sitter space. That such a decoupling occurs in an expanding background is a highly nontrivial result.

The results obtained here in the large-$N$ limit are new exact results in the context of QFT on de Sitter. Since the early work of Starobinsky and Yokoyama \cite{Starobinsky:1994bd}, nonperturbative calculations have been rather scarce, owing to the difficulty of extending  to de Sitter space standard flat space-time resummation or nonperturbative techniques. The situation has improved in recent years with, in particular, the use of the dynamical renormalization group approach \cite{Burgess:2009bs}, explicit QFT calculations in various nonperturbative approximation schemes \cite{Riotto:2008mv,Serreau:2011fu,Prokopec:2011ms,Arai:2011dd,Boyanovsky:2012qs}, all orders perturbative results \cite{Marolf:2010nz,Hollands:2010pr,Korai:2012fi}, and resummation of Schwinger-Dyson equations \cite{Garbrecht:2011gu,Jatkar:2011ju,Akhmedov:2011pj,Youssef:2013by}. Also worth mentioning is the interesting proposal for a reorganized perturbative expansion in the context of Euclidean de Sitter space proposed in \cite{Rajaraman:2010xd,Beneke:2012kn}. 

We believe the methods proposed here and in \cite{Parentani:2012tx} provide a powerful tool to pursue these studies. In particular, they are well suited to the study of the IR properties of de Sitter correlators and are in principle applicable to other approximation schemes. Examples presently under study include $1/N$ corrections to the two-point function in the $O(N)$ model and approximations based on two-particle-irreducible techniques \cite{Gautier}.

\section*{Acknowledgements}

We acknowledge useful discussions with X. Busch, M. Garny, F. Gautier and U. Reinosa.

\appendix

\section{Proof of \Eqn{eq:IF2}}
\label{appsec:proof}

Consider the integral on the right-hand side of \Eqn{eq:IF2} and use \Eqn{eq:IF} to eliminate $\hat\Pi_H$:
\bea
\int_{p}^\infty \!\!ds \,\hat I_\rho(p,s )\hat \Pi_H(s ,p')=\int_{p}^\infty \!\!ds \,\hat I_\rho(p,s )\hat I_F(s ,p')&&\nn
\label{appeq:secondline}
-\int_{p}^\infty \!\!ds \int_s^\infty\!\!ds'\,\hat I_\rho(p,s )\hat \Pi_\rho(s ,s')\hat I_F(s',p').&&
\eea
Writing the double integral on the right-hand side as
\beq
 \int_{p}^\infty \!\!ds \int_s^\infty\!\!ds'=\int_{p}^\infty \!\!ds' \int_p^{s'}\!\!ds
\eeq
and using \Eqn{eq:Irho}, the second line of \Eqn{appeq:secondline} rewrites as
\bea
&&-\int_{p}^\infty \!\!ds' \int_p^{s'}\!\!ds\,\hat I_\rho(p,s )\hat \Pi_\rho(s ,s')\hat I_F(s',p')\nn
&&=\int_{p}^\infty \!\!ds' \left\{\hat \Pi_\rho(p,s')-\hat I_\rho(p,s')\right\}\hat I_F(s',p').
\eea
We thus get
\beq
  \int_{p}^\infty \!\!ds \,\hat I_\rho(p,s )\hat \Pi_H(s ,p')=\int_{p}^\infty \!\!ds\,\hat \Pi_\rho(p,s)\hat I_F(s,p'),
\eeq
hence \Eqn{eq:IF2}.

\section{Two-point correlators}
\label{appsec:tpt}

In this series of appendices, we discuss various technical aspects of the present work. Appendices \ref{appsec:tpt} and \ref{appsec:piIR} present material directly used in the main body of the paper, in particular the calculation of the one-loop bubble functions $\hat\Pi_F^{\rm IR}$ and $\hat\Pi_\rho^{\rm IR}$, Eqs. \eqn{eq:PiFIR}-\eqn{eq:PirhoIR}. The remaining Appendices \ref{appsec:pimix}--\ref{appsec:renorm} provide a detailed discussion of the role of high momenta (Appendix \ref{appsec:roleofUV}), including the question of renormalization (Appendix \ref{appsec:renorm}), in the case $D=d+1=4$, where the theory \eqn{eq:theory} is perturbatively renormalizable. This requires the evaluation of the functions  $\hat\Pi_{F,\rho}^{\rm mix}$ and  $\hat\Pi_{F,\rho}^{\rm UV}$, defined below and discussed in Appendices \ref{appsec:pimix} and \ref{appsec:piUV} respectively.

In the present section, we recall the form of the statistical and spectral two-point functions and their approximate forms in the IR and UV limits. As in the main text, we separate the two regimes with an arbitrary scale $\mu\sim1$. For a generic two-point function ${\cal D}(p,p')$, we introduce the notations
\bea
\label{appeq:IRdef}
 {\cal D}^{\rm IR}(p,p')&=&{\cal D}(p\lesssim\mu,p'\lesssim\mu),\\
\label{appeq:UVdef}
 {\cal D}^{\rm UV}(p,p')&=&{\cal D}(p\gtrsim\mu,p'\gtrsim\mu),
\eea
for the cases where both momenta are IR or UV and
\bea
\label{appeq:mixdef}
 {\cal D}^{\rm mix}(p,p')&=&{\cal D}(p\lesssim\mu,p'\gtrsim\mu),\\
\label{appeq:ximdef}
 {\cal D}^{\rm xim}(p,p')&=&{\cal D}(p\gtrsim\mu,p'\lesssim\mu),
\eea
for the cases where one momentum in IR and the other is UV. Notice that for (anti)symmetric functions ${\cal D}(p,p')=\pm {\cal D}(p',p)$, one has ${\cal D}^{\rm mix}(p,p')=\pm {\cal D}^{\rm xim}(p',p)$.

The statistical and spectral two-point functions of a free scalar field of mass $M$ in the Bunch-Davies vacuum are given by
\bea
 \hat F(p,p')&=&{\pi\over4}\sqrt{pp'}\,{\rm Re}\left\{H_\nu(p)H^*_\nu(p')\right\},\nn
 \hat\rho(p,p')&=&-{\pi\over2}\sqrt{pp'}\,{\rm Im}\left\{H_\nu(p)H^*_\nu(p')\right\},
\eea
with $\nu=\sqrt{d^2/4-M^2}$. In the next sections we shall use the propagator \eqn{eq:G} with both $p$ and $p'$ on the upper branch $\hat\C^+$ of the contour $\hat\C$, where ${\rm sign}_{\hat\C}(p-p')=-{\rm sign}(p-p')$ and thus
\beq
\label{appeq:propag}
 \hat G(p,p')=\hat F(p,p')+{i\over2}{\rm sign}(p-p')\hat\rho(p,p')\,\,\,{\rm for}\,\,\,p,p'\in\hat\C^+.
\eeq
We recall the symmetry relations $\hat G(p,p')=\hat G(p',p)$, $\hat F(p,p')=\hat F(p',p)$ and $\hat \rho(p,p')=-\hat \rho(p',p)$.

We are concerned here with the small mass case \hbox{$M\lesssim 1$} for which the index $\nu$ is real. The leading low and high momentum behaviors of the Hankel function are given by \cite{Gradshteyn}
\beq
\label{appeq:HnuIR}
 H_\nu(p)\sim \frac{\Gamma(\nu)}{i\pi}\left(2\over p\right)^{\!\nu}\equiv-i\sqrt{4\over\pi p}f_\nu(p),\,\,{\rm for}\quad p\ll1,
\eeq
where the second relation defines the function $f_\nu(p)$, and
\beq
\label{appeq:HnuUV}
 H_\nu(p)\sim \sqrt{2\over\pi p}\,e^{i\left(p-\varphi_\nu\right)}, \,\,{\rm for}\quad p\gg1,
\eeq
where $\varphi_\nu={\pi\over2}(\nu+{1\over2})$.

The IR behavior of the statistical function is readily obtained as
\beq
\label{appeq:FIR}
 \hat F_{\rm IR}(p,p')=f_\nu(p)f_\nu(p')=\frac{F_\nu}{(pp')^{\nu-{1\over2}}},
\eeq
where $F_\nu=[2^\nu\Gamma(\nu)]^2/4\pi$. The factorization property \eqn{appeq:FIR} expresses the (approximate~\cite{Campo:2005sy}) classical nature and the phase coherence of IR fluctuations in de Sitter \cite{Guth:1985ya,Campo:2003pa}. The leading IR behavior of the spectral function necessitates a more precise calculation. It can be written as
\bea
\label{appeq:rhoIR}
 \hat \rho_{\rm IR}(p,p')&=&-\frac{\sqrt{pp'}}{2\nu}\left[\left({p\over p'}\right)^{\!\!\nu}\!-\left({p'\over p}\right)^{\!\!\nu}\right]\nn
 &=&-\sqrt{pp'}\,{\cal P}_\nu\left(\ln{p\over p'}\right),
\eea
where we introduced the function ${\cal P}_\nu$ of \eqn{eq:P}.
The mixed components are obtained from \eqn{appeq:HnuIR}, \eqn{appeq:HnuUV}
\bea
\label{appeq:Fmix}
 \hat F_{\rm mix}(p,p')&=&-f_\nu(p)\,\frac{\sin(p'-\varphi_\nu)}{\sqrt 2},\\
\label{appeq:rhomix}
 \hat \rho_{\rm mix}(p,p')&=&2f_\nu(p)\,\frac{\cos(p'-\varphi_\nu)}{\sqrt 2},
\eea
or, equivalently, 
\beq
\label{appeq:Gmix}
 \hat G_{\rm mix}(p,p')=f_\nu(p)\frac{e^{-i(p'-\varphi_\nu)}}{i\sqrt 2},\quad{\rm for}\quad p,p'\in\hat\C^+.
\eeq
Finally, the UV components have the following Minkowski-like expressions (with physical momentum in place of time)
\bea
\label{appeq:FUV}
 \hat F_{\rm UV}(p,p')&=&{1\over2}\cos(p-p'),\\
\label{appeq:rhoUV}
 \hat \rho_{\rm UV}(p,p')&=&-\sin(p-p'),
\eea
or, equivalently,
\beq
\label{appeq:GUV}
 \hat G_{\rm UV}(p,p')=\frac{e^{-i|p-p'|}}{2},\quad{\rm for}\quad p,p'\in\hat\C^+.
\eeq
In the deep UV regime $p,p'\gg1$ and for subhorizon time separation $|\ln p/p'|=|t-t'|\ll1$, one has $p-p'\approx -p(t-t')$ (recall that  $p'e^{t'}=p\,e^{t}$) and one recovers the (massless) Minkowski vacuum correlator
\beq
 \frac{\hat G_{\rm UV}(p,p')}{\sqrt{pp'}}\approx \frac{e^{-ip|t-t'|}}{2p}.
\eeq

\section{Calculation of $\hat\Pi^{\rm IR}(p,p')$}
\label{appsec:piIR}

Here, we compute the leading IR behavior of the one-loop bubble function $\hat\Pi(p,p')$, defined in \eqn{eq:Np1}. It is sufficient to consider the case where both $p$ and $p'$ lie on the upper branch $\hat\C^+$ of the contour and $p<p'$, such that ${\rm sign}_{\hat\C}(p-p')=1$. We separate the momentum integral in three regions: low momentum modes, such that $qp<qp'\lesssim\mu$;  intermediate momentum modes with $qp\lesssim\mu\lesssim qp'$; and high momentum modes with $\mu\lesssim qp<qp'$.

\subsection{IR contribution}

The contribution from low momentum, IR modes gives the dominant contribution. One needs to consider the $F$ and $\rho$ components, Eqs. \eqn{eq:PiF} and \eqn{eq:Pirho} separately. Here, the condition $qp<qp'\lesssim\mu$ implies that $rp<rp'\lesssim\mu$ and both propagators under the loop integral can be approximated by their IR behavior, Eqs. \eqn{appeq:FIR} and \eqn{appeq:rhoIR}. In that case, due to the strong IR enhancement of the statistical propagator as compared to the spectral one, one can safely neglect the (quantum) contribution $\hat\rho\hat\rho$ as compared to the (classical) one\footnote{This is a typical classical (statistical) field approximation \cite{Martin:1973zz,Aarts:1997kp,vanderMeulen:2007ah}.} $\hat F\hat F$ in $\hat\Pi_F$. We get
\bea
 \hat\Pi_F^{\rm IR}(p,p')\!&\approx&\!-\frac{\lambda}{6}(pp')^{d-3\over2}\!\!\int_{|\bq|<{\mu\over p'}}\!\!\frac{\hat F_{\rm IR}(qp,qp')\hat F_{\rm IR}(rp,rp')}{qr}\nn
 \label{appeq:PiFIRint}
 &\approx&\!-\frac{\lambda}{6}(pp')^{d-3\over2}\hat F^2_{\rm IR}(p,p')\int_\bq\frac{1}{(qr)^{2\nu}},
\eea
for the $F$ component and
\bea
 \hat\Pi_\rho^{\rm IR}(p,p')\!&\approx&\!-\frac{\lambda}{3}(pp')^{d-3\over2}\!\!\int_{|\bq|<{\mu\over p'}}\!\!\frac{\hat F_{\rm IR}(qp,qp')\hat \rho_{\rm IR}(rp,rp')}{qr}\nn
 \label{appeq:PirhoIRint}
 &&\hspace{-1.3cm}=-\frac{\lambda}{3}(pp')^{d-3\over2}\hat F_{\rm IR}(p,p')\hat \rho_{\rm IR}(p,p')\int_{|\bq|<{\mu\over p'}}\frac{1}{q^{2\nu}},
\eea
for the $\rho$ component, where we recall that $r=|\be+\bq|$ with $\be$ an arbitrary unit vector. Since the momentum integral in the second line of \Eqn{appeq:PiFIRint} converges rapidly at large momentum ($2\nu\approx d$) we can safely ignore the upper bound. One cannot do so in the momentum integral in the second line of \Eqn{appeq:PirhoIRint} since the latter is sensitive to the upper bound. This results in a nontrivial  $p'/p$ dependence, as discussed below. 

The integral in \Eqn{appeq:PiFIRint} is easily evaluated by using the technique of Feynman parameters. We write
\beq
 \frac{1}{(qr)^{2\nu}} =\frac{\Gamma(2\nu)}{\Gamma^2(\nu)}\int_0^1dx\,\frac{[x(1-x)]^{\nu-1}}{\left[xq^2+(1-x)r^2\right]^{2\nu}}
\eeq
and $xq^2+(1-x)r^2=[\bq+(1-x)\be]^2+x(1-x)$. Performing the shift $\bq+(1-x)\be\to\bq$ under the $\bq$ integration, one can perform all the angular integrations in $d$-dimensional spherical coordinates, to get
\beq
 \int_\bq\frac{1}{(qr)^{2\nu}}=\frac{\Omega_d}{(2\pi)^d}\frac{\Gamma(2\nu)}{\Gamma^2(\nu)}\int_0^1\!\!dx\!\int_0^\infty\!\!dq\,\frac{q^{d-1}[x(1-x)]^{\nu-1}}{\left[q^2+x(1-x)\right]^{2\nu}},
\eeq
where $\Omega_d=2\pi^{d/2}/\Gamma(d/2)$.
Finally, performing the change of variable $u=q^2/x(1-x)$, the $u$ and $x$ integration can be trivially performed. We obtain, recalling $\nu=d/2-\varepsilon$, 
\bea
 \int_\bq\frac{1}{(qr)^{2\nu}}&=&\frac{\Omega_d}{(2\pi)^d}\frac{\Gamma\left({d\over2}\right)\Gamma\left({d\over2}-2\varepsilon\right)}{\Gamma^2\left({d\over2}-\varepsilon\right)}\frac{\Gamma^2(\varepsilon)}{2\Gamma(2\varepsilon)}\nn
 \label{appeq:IRint2}
 &=&\frac{\Omega_d}{(2\pi)^d}\frac{1}{\varepsilon}\Big[1+{\cal O}(\varepsilon)\Big].
\eea
The leading IR behavior is then
\beq
 \hat\Pi_F^{\rm IR}(p,p')=\frac{\pi_F}{(pp')^{\kappa+{1\over2}}}
\eeq
with $\kappa={d\over2}-2\varepsilon=\nu-\varepsilon<\nu$ and 
\beq
 \pi_F=-\frac{\lambda F^2_\nu}{6\varepsilon}\frac{\Omega_d}{(2\pi)^d}.
\eeq 
As already emphasized below  \Eqn{eq:factorization}, the function $\hat\Pi_F^{\rm IR}$ assumes a factorized form which is reminiscent of the classical coherent nature of IR de Sitter fluctuations. For later use we rewrite $\hat\Pi_F^{\rm IR}$ as
\beq
\label{appeq:factorized}
 \hat\Pi_F^{\rm IR}(p,p')=A(p)A(p'),
\eeq
where the function $A$ is related to the one introduced in \eqn{eq:Adef} as 
\beq
 A(p)=\frac{i\sqrt{-\pi_F}}{p^{\kappa+{1\over2}}}=i\sqrt{-\pi_F\over p}\,{\cal A}\left(\ln {p\over\mu}\right).
\eeq

The integral in \Eqn{appeq:PirhoIRint} is easily performed:
\bea
 \int_{|\bq|<{\mu\over p'}}\frac{1}{q^{2\nu}}&=&\frac{\Omega_d}{(2\pi)^d}\frac{1}{2\varepsilon}\left(\mu\over p'\right)^{\!2\varepsilon}\nn
 \label{appeq:IRint}
 &=&\frac{\Omega_d}{(2\pi)^d}\frac{(p')^{-2\varepsilon}}{2\varepsilon}\Big[1+{\cal O}(\varepsilon)\Big],
\eea
which leads to
\beq
 \hat\Pi_\rho^{\rm IR}(p,p')=\frac{\pi_\rho}{\sqrt{pp'}}\,\left(p\over p'\right)^{\!\varepsilon}\!{\cal P}\left(\ln{p\over p'}\right)\quad{\rm for}\quad p<p',
\eeq
where 
\beq
 \pi_\rho=\frac{\lambda F_\nu}{6\varepsilon}\frac{\Omega_d}{(2\pi)^d}.
\eeq
Using the antisymmetry of the function $\hat\Pi_\rho(p,p')$, we have then
\beq
 \hat\Pi_\rho^{\rm IR}(p,p')=\frac{\pi_\rho}{\sqrt{pp'}}\,\left(p'\over p\right)^{\!\varepsilon}\!{\cal P}\left(\ln{p\over p'}\right)\quad{\rm for}\quad p>p'.
\eeq
From this we get \Eqn{eq:PirhoIR}.

\subsection{Mixed and UV contributions}

We now show that the intermediate and high momentum contributions to $\hat\Pi^{\rm IR}$ are subleading in the limit $p,p'\ll1$.  The contribution from intermediate momenta $qp\lesssim\mu\lesssim qp'$ is such that $q\gtrsim\mu/p'\gg1$, which implies that $r\approx q\gg1$ and thus $rp\lesssim1\lesssim rp'$. It follows that both propagators in \Eqn{eq:Np1} can be replaced by their ``mixed'' expression \eqn{appeq:Gmix}. We thus have to evaluate the integral
\bea
 &&\int_{{\mu\over p'}< |\bq|<{\mu\over p}}{G_{\rm mix}(qp,qp')\hat G_{\rm mix}(rp,rp')\over qr}\nn
 &=&\frac{F_{\rm IR}(p,p)}{2}\int_{{\mu\over p'}<|\bq|<{\mu\over p}}\frac{e^{-i\left[(q+r)p'-2\varphi_\nu\right]}}{(qr)^{\nu+{1\over2}}}\nn
 &\approx&\frac{\Omega_d}{(2\pi)^d}\frac{F_{\rm IR}(p,p)}{2}\int_{\mu\over p'}^{\mu\over p}\frac{dq}{q^{2}}e^{-2i(qp'-\varphi_{d/2})},
\eea
where we can safely replace $\nu\to d/2$ in the last line. The remaining integral is bounded:
\beq
 \left|\int_{\mu\over p'}^{\mu\over p}\frac{dq}{q^{2}}e^{-2i(qp'-\varphi_\nu)}\right|<\int_{\mu\over p'}^{\mu\over p}\frac{dq}{q^{2}}=\frac{p'-p}{\mu}
\eeq
and one easily checks that the corresponding contributions to either $\hat\Pi_F^{\rm IR}={\rm Re}\,\hat\Pi^{\rm IR}$ or $ \hat\Pi_\rho^{\rm IR}=-2i{\rm Im}\,\hat\Pi^{\rm IR}$ are negligible as compared to those obtained in the previous subsection (notice also that there is no IR enhancement factor $1/\varepsilon$).

Finally, the high momentum contribution can be estimated along similar lines. In $D\ge4$ the integral diverges for $p'\to p$ and needs to be regulated. Here, we regulate the theory by imposing a sharp cutoff $\Lambda$ on physical momenta, i.e. by effectively replacing the propagators $G(p,p')\to G(p,p')\theta(\Lambda-p)\theta(\Lambda-p')$, see Appendix \ref{appsec:renorm}. We thus have $\mu\lesssim qp<qp'<\Lambda$ and, therefore, $q\gg1$, so that $r\approx q\gg1$ and $rp'>rp\gtrsim\mu$. Both propagators in \eqn{eq:Np1} can thus be replaced by their UV expression \eqn{appeq:GUV}. The corresponding contribution reads
\bea
 &&\int_{{\mu\over p}<  |\bq|<{\Lambda\over p}}{\hat G_{\rm UV}(qp,qp')\hat G_{\rm UV}(rp,rp')\over qr}\nn
 &&\qquad=\frac{1}{4}\int_{{\mu\over p}<  |\bq|<{\Lambda\over p}}\frac{e^{i(q+r)[p-p']}}{qr}\nn
\label{appeq:UVcontrib}
 &&\qquad\approx\frac{\Omega_d}{4(2\pi)^d}\int_{\mu\over p}^{\Lambda\over p}\frac{dq}{q^{3-d}}e^{2iq(p-p')}.
\eea
Again an upper bound is given by
\beq
 \left|\int_{\mu\over p}^{\Lambda\over p}\frac{dq}{q^{3-d}}e^{2iq(p-p')}\right|<\int_{\mu\over p}^{\Lambda\over p}\frac{dq}{q^{3-d}}=\frac{\left({\Lambda\over p}\right)^{\!d-2}\!\!-\left({\mu\over p}\right)^{\!d-2}}{d-2},
\eeq
which is easily shown to give negligible contributions to both $\hat\Pi_F^{\rm IR}$ and $ \hat\Pi_\rho^{\rm IR}$ in the low momentum limit $p,p'\ll1$, at fixed cutoff $\Lambda$.

We end this subsection with a more precise evaluation of the UV contribution \eqn{appeq:UVcontrib}, useful for our discussion of renormalization in Appendix \ref{appsec:renorm}. One has, for $p\ll1$ and $x=p-p'$ finite,
\beq
\label{appeq:UVcontrib2}
 \int_{\mu\over p}^{\Lambda\over p}\frac{dq}{q^{3-d}}e^{2iqx}\approx\left({p\over\Lambda}\right)^{3-d}\frac{e^{2i{\Lambda\over p}x}}{2ix}-\left({p\over\mu}\right)^{3-d}\frac{e^{2i{\mu\over p}x}}{2ix}.
\eeq
We see that, as expected, that the one-loop bubble---which we recall is a one-loop contribution to the four-point vertex function---is finite in the limit $\Lambda\to\infty$ for $D=d+1\le3$, where the present theory is super-renormalizable, and has no coupling divergence. The case $D=4$, where the theory is (perturbatively) renormalizable, requires a more careful analysis, which is performed in Sec. \ref{appsec:renorm}. For a larger number of dimensions, the one-loop bubble exhibits a strong cutoff dependence.

\section{Calculation of $\hat\Pi^{\rm mix}(p,p')$}
\label{appsec:pimix}

For our discussion of the role of high momentum (subhorizon) modes to the integral equations \eqn{eq:Irho}-\eqn{eq:PiH} in Appendix \ref{appsec:roleofUV} below, we shall need both the mixed ($p\lesssim\mu\lesssim p'$) and UV ($\mu\lesssim p,p'$) behaviors of the one-loop bubble $\hat\Pi(p,p')$ for large $p'$. We analyze the former in detail here, specializing to $d=3$ spatial dimensions. 

By definition, see \Eqn{appeq:mixdef}, the mixed contribution $\hat\Pi^{\rm mix}(p,p')$ is such that $p<p'$. The propagator \eqn{appeq:propag} thus reads
\beq
 \hat G(p,p')={\pi\over4}\sqrt{pp'}H_\nu(p)H_\nu^*(p').
\eeq

By choosing $q=|\bq|$ and $r=|\be+\bq|$ as integration variables, the loop integral \eqn{eq:Np1} can be written as
\beq
 \hat \Pi(p,p')=-{\lambda\over24\pi^2}J(p,p'),
\eeq
with
\bea
 &&\hspace{-.5cm}J(p,p')=\int_0^\infty dq\,\hat G(qp,qp')\int_{|1-q|}^{1+q}dr\,\hat G(rp,rp')\nn
\label{appeq:Jdef}
 &&\hspace{-.5cm}=\int_0^\infty dq\,{\cal H}(q,p,p')\Big\{{\cal F}(1+q,p,p')-{\cal F}(|1-q|,p,p')\Big\}.\nn
\eea
In the second line we introduced the (redundant) notation
\beq
 {\cal H}(q,p,p')=\hat G(qp,qp')
\eeq
for later use. We exploit the fact that, in $d=3$, the $r$ integration can be performed explicitly using the indefinite integral \cite{Gradshteyn,Boyanovsky:2012qs}
\bea
 &&{\cal F}(q,p,p')=\int dq \,\hat G(qp,qp')=\nn
 &&={\pi\over4}\sqrt{pp'}\,\frac{qp'H_\nu(qp)H_{\nu-1}^*(qp')-qpH_\nu^*(qp')H_{\nu-1}(qp)}{p^2-p^{\prime2}}\nn
\label{appeq:function}
 &&=-\hat G(qp.qp')\frac{{\cal R}_\nu(qp)-{\cal R}_\nu^*(qp')}{q(p^2-p^{\prime2})},
\eea
where we defined the function
\beq
 {\cal R}_\nu(p)=\frac{pH_{\nu-1}(p)}{H_\nu(p)}.
\eeq
The third line of \Eqn{appeq:function} is particularly useful when it comes to approximating the function ${\cal F}(q,p,p')$ for low or large momenta. Indeed, the function ${\cal R}_\nu(p)$ has the following simple limits:
\beq
 {\cal R}_\nu(p\gg1)\approx ip\quad {\rm and}\quad{\cal R}_\nu(p\ll1)\approx\frac{p^2}{2(\nu-1)}.
\eeq

We are interested in the behavior of the function $J(p,p')$ for $p\ll1\ll p'$. In the vanishing mass limit, $\nu\to3/2$, the integral features IR divergences for $qp'\lesssim 1$ and $|1-q|p'\lesssim 1$ as can be seen from the low momentum behavior of the integrand, see e.g., Eqs. \eqn{appeq:HIR} and \eqn{appeq:FfIR} below. Using a similar technique as in \cite{Boyanovsky:2012qs, vanderMeulen:2007ah}, we separate the $q$ integral in \eqn{appeq:Jdef} in four contributions\footnote{Note that the integral is UV finite for $p\neq p'$, see Appendix \ref{appsec:renorm}. We can safely send the upper bound to $\infty$.}
\beq
 \int_0^\infty=\int_0^{\mu\over p'}+\int_{\mu\over p'}^{1-{\mu\over p'}}+\int_{1-{\mu\over p'}}^{1+{\mu\over p'}}+\int_{1+{\mu\over p'}}^\infty,
 \eeq
with $\mu\sim1$.
Reorganizing the various terms and using some changes of variables, we arrive at
\beq
J(p,p')=\sum_{i=1}^4J_i(p,p'),
\eeq
where (omitting the implicit dependence on $p$ and $p'$ for brevity)
\bea
 J_1&=&\int_0^{\mu\over p'}\!\!dq\,{\cal H}(q)\{{\cal F}(1+q)-{\cal F}(1-q)\},\\
 J_2&=&-\int_0^{\mu\over p'}\!\!dq\,{\cal F}(q)\{{\cal H}(1+q)+{\cal H}(1-q)\}
\eea
grab together all potentially IR dangerous contributions,
\beq
 J_3=\int_{\mu\over p'}^\infty\!\!\!dq\,\{{\cal H}(q){\cal F}(1+q)-{\cal H}(1+q){\cal F}(q)\}
\eeq
and
\beq
 J_4=-\int_{\mu\over p'}^{1-{\mu\over p'}}\!\!\!dq\,{\cal H}(q){\cal F}(1-q).
\eeq

In the following, we adopt a similar notation as the one introduced in Appendix \ref{appsec:tpt} for two-point functions to distinguish the various regimes in $q$ of the functions ${\cal H}$ and ${\cal F}$. For instance we note ${\cal H}(q,p,p')$ as
\bea
 {\cal H}_{\rm IR}(q,p,p')\quad &{\rm for}&\quad qp,qp'\lesssim\mu,\\
 {\cal H}_{\rm mix}(q,p,p')\quad &{\rm for}&\quad qp\lesssim \mu\lesssim qp',\\
 {\cal H}_{\rm UV}(q,p,p')\quad &{\rm for}&\quad \mu\lesssim qp,qp',
\eea 
and similarly for ${\cal F}(q,p,p')$.
With this notation, we have, for the IR contributions,
\beq
 J_1\approx\int_0^{\mu\over p'}\!\!dq\,{\cal H}_{\rm IR}(q)\{{\cal F}_{\rm mix}(1+q)-{\cal F}_{\rm mix}(1-q)\}
\eeq
and
\beq
 J_2\approx-\int_0^{\mu\over p'}\!\!dq\,{\cal F}_{\rm IR}(q)\{{\cal H}_{\rm mix}(1+q)+{\cal H}_{\rm mix}(1-q)\}.
\eeq
We separate the integral $J_3$ in two contributions from intermediate $qp\lesssim \mu\lesssim qp'$ and high $\mu\lesssim qp,qp'$ momenta:
\beq
 J_3=J_3^{\rm mix}+J_3^{\rm UV},
\eeq
with
\beq
 J_3^{\rm mix}\approx\int_{\mu\over p'}^{\mu\over p}\!\!\!dq\,\{{\cal H}_{\rm mix}(q){\cal F}_{\rm mix}(1+q)-{\cal H}_{\rm mix}(1+q){\cal F}_{\rm mix}(q)\}
\eeq
and
\beq
\label{appeq:J3UV}
 J_3^{\rm UV}\approx\int_{\mu\over p}^\infty\!\!\!dq\,\{{\cal H}_{\rm UV}(q){\cal F}_{\rm UV}(1+q)-{\cal H}_{\rm UV}(1+q){\cal F}_{\rm UV}(q)\}.
\eeq
Finally
\beq
 J_4\approx-\int_{\mu\over p'}^{1-{\mu\over p'}}\!\!\!dq\,{\cal H}_{\rm mix}(q){\cal F}_{\rm mix}(1-q).
\eeq

To evaluate the various integrals above, we use the leading IR and mix behaviors
\bea
\label{appeq:HIR}
 {\cal H}_{\rm IR}(q,p,p')&\approx&\frac{\hat G_{\rm IR}(p,p')}{q^{2\nu-1}},\\
\label{appeq:FfIR}
 {\cal F}_{\rm IR}(q,p,p')&\approx&-q{\cal H}_{\rm IR}(q,p,p')
\eea
and
\bea
 {\cal H}_{\rm mix}(q,p,p')&\approx&\sqrt{\frac{\hat G_{\rm IR}(p,p)}{2}}\frac{e^{-i(qp'-\varphi_\nu)}}{iq^{\nu-{1\over2}}},\\
 {\cal F}_{\rm mix}(q,p,p')&\approx&{\cal H}_{\rm mix}(q,p,p')\left(\frac{i}{p'}+q\frac{p^2}{p^{\prime2}}\right).
\eea
The UV contribution \eqn{appeq:J3UV} has no IR problem and can be safely evaluated by setting $\nu\to3/2$. Using
\beq
 H_{1\over2}(p)=\sqrt{2\over\pi p}{e^{ip}\over i}\,,\quad H_{3\over2}(p)=H_{1\over2}(p)\left({1\over p}-i\right),
\eeq
we get, after some calculations, 
\bea
\label{appeq:UV1}
 &&\hspace{-1cm}{\cal H}_{\rm UV}(q)=\frac{e^{iq(p-p')}}{2}\left[1+{i\over q}\left({1\over p}-{1\over p'}\right)+{1\over qp p'}\right],\\
\label{appeq:UV2}
 &&\hspace{-1cm}{\cal F}_{\rm UV}(q)=\frac{e^{iq(p-p')}}{2i(p-p')}\left[1+{i\over q}\left({1\over p}-{1\over p'}\right)\right].
\eea
We thus have (recall that $qp,qp'\gg1$)
\bea
 &&{\cal H}_{\rm UV}(q){\cal F}_{\rm UV}(1+q)-{\cal H}_{\rm UV}(1+q){\cal F}_{\rm UV}(q)\nn
 &&=\frac{ie^{i(2q+1)(p-p')}}{4pp'(p'-p)}\frac{2q+1+i\left({1\over p}-{1\over p'}\right)}{\left[q(q+1)\right]^2}\nn
\label{appeq:UV3}
 &&\approx\frac{ie^{-i(2q+1)p'}}{4pp^{\prime 2}}\frac{2}{q^3}.
\eea
Notice that the high momentum $q^{-3}$ behavior shows that, as expected, the integral $J(p,p')$ is UV finite for $p\neq p'$.

Using the various approximations above in the relevant integrals and isolating the dominant contributions in the limit $p\ll1\ll p'$, we get, after some manipulations,
\bea
 J_1+J_2&\approx&i\hat G_{\rm IR}(p,p)\frac{e^{-ip'}}{\left(p'\right)^{1+\epsilon}}\int_0^\mu dx\frac{\sin x+x\cos x}{x^{1-2\varepsilon}}\nn
 &\approx&i\frac{\hat G_{\rm IR}(p,p)}{\varepsilon}\frac{e^{-ip'}}{\left(p'\right)^{1+\epsilon}}
\eea
and
\bea
 J_4&\approx&i\hat G_{\rm IR}(p,p)\frac{e^{-ip'}}{2p'}\int_{\mu\over p'}^{1-{\mu\over p'}}\frac{dq}{\left[q(1-q)\right]^{1-\varepsilon}}\nn
 &\approx& i\frac{\hat G_{\rm IR}(p,p)}{\varepsilon}\frac{e^{-ip'}}{p'}\left[1-\left({1\over p^{\prime}}\right)^{\!\!\varepsilon}\,\right].
\eea
The contributions $J_3^{\rm mix}$ and $J_3^{\rm UV}$ are suppressed by relative factors $\varepsilon p^2/p'$ and $\varepsilon p^4/p^{\prime2}$ respectively. The IR contribution $J_1+J_2$ cancels the $1/(p')^\varepsilon$ contribution from $J_4$ and we obtain, finally,
\beq
 J(p,p')\approx i\frac{\hat G_{\rm IR}(p,p)}{\varepsilon}\frac{e^{-ip'}}{p'}\,.
\eeq
We conclude that the mixed one-loop bubble $\hat\Pi^{\rm mix}$ assumes the factorized form
\beq
\label{appeq:Pimixfac}
 \hat \Pi^{\rm mix}(p,p')\approx A(p)B(p'),
\eeq
where the low momentum piece $A(p)$ is the one characterizing the IR contribution \eqn{appeq:factorized} with, here, $\kappa+1/2=2-2\varepsilon$ and $\pi_F=-\lambda/48\pi^2\varepsilon$, and the high momentum piece is given by
\beq
\label{appeq:B}
 B(p)=-\sqrt{-\pi_F}\frac{e^{-ip}}{p}.
\eeq

\section{Calculation of $\hat\Pi^{\rm UV}(p,p')$}
\label{appsec:piUV}

We shall also need the high momentum behavior of the one-loop bubble $\hat\Pi^{\rm UV}$. We proceed along similar lines as in the previous section. As in the case of $\hat\Pi^{\rm IR}$ discussed in Appendix \ref{appsec:piIR}, one needs to regulate the integral in the UV since it diverges for $p'\to p$. We employ the same regulator as before, namely a sharp cutoff on physical momenta. This essentially\footnote{In principle, one should consider the fact that the cutoff also applies to $r=|\be+\bq|$: there should be a term $\theta(\Lambda-qp)\theta(\Lambda-qp')\theta(\Lambda-rp)\theta(\Lambda-rp')$ under the $\bq$ integral. However, the UV divergence arises from very high momenta where $r\approx q$ and we neglect these subtleties here.} amounts to cutting the $q$ integration with a sharp cutoff $\Lambda/\max(p,p')$.

We shall see that $\hat\Pi^{\rm UV}(p,p')$ has a singular $1/|p-p'|$ structure when $p'\to p$. For our present purposes, this function is essentially involved under integrals such as \eqn{eq:Np2} which, we assume, are dominated by this singular structure. Therefore, we seek the leading singular contribution to $\hat\Pi^{\rm UV}(p,p')$.

In the present case, where $p,p'\gtrsim 1$, potential IR contributions to the momentum integral in \eqn{appeq:Jdef} come from the regions (taking $p<p'$) $q\lesssim\mu/p'$ and $|1-q|\lesssim\mu/p'$ as well as $\mu/p'\lesssim q\lesssim\mu/p$ and $\mu/p'\lesssim |1-q|\lesssim\mu/p$. However, these give regular contributions in the limit $p'\to p$. The remaining contributions can be separated in three pieces:
\beq
 J^{\rm UV}_1=-\int_{\mu\over p}^{1-{\mu\over p}}dq \,{\cal H}_{\rm UV}(q){\cal F}_{\rm UV}(1-q),
\eeq
\beq
J^{\rm UV}_2 =\int_{\mu\over p}^{{\Lambda\over p}}dq\,\{{\cal H}_{\rm UV}(q){\cal F}_{\rm UV}(1+q)-{\cal H}_{\rm UV}(1+q){\cal F}_{\rm UV}(q)\}
\eeq
and
\beq
 J^{\rm UV}_3=\int_{{\Lambda\over p}-1}^{{\Lambda\over p}}dq \,{\cal H}_{\rm UV}(1+q){\cal F}_{\rm UV}(q),
\eeq
which can be safely evaluated by setting $\nu\to 3/2$, i.e., using Eqs. \eqn{appeq:UV1}-\eqn{appeq:UV3}. We obtain, for the leading singular behavior when $p'\to p$ (we also take ${\Lambda\over p}\gg1$),
\bea
 J^{\rm UV}_1&\approx&\frac{e^{i(p-p')}}{4i(p-p')},\\
 J^{\rm UV}_3&\approx&\frac{e^{2i{\Lambda\over p}(p-p')}}{4i(p-p')}\frac{\sin(p-p')}{p-p'},
\eea
whereas $J^{\rm UV}_2\approx  J^{\rm UV}_1/p$ receives a $1/p$ suppression. 

Considering also the case $p>p'$, our final result for the leading singular behavior of the function $\hat\Pi^{\rm UV}$ is
\beq
\label{appeq:PiUV}
 \hat\Pi^{\rm UV}(p,p')=\frac{\lambda}{96i\pi^2}\left\{\frac{e^{-i|x|}}{|x|}+\frac{e^{-2i\tilde\Lambda |x|}}{|x|}\frac{\sin x}{x}\right\},
\eeq
where $x\equiv p-p'$ and $\tilde\Lambda=\Lambda/\max(p,p')$.

As in the case of the field correlator \eqn{appeq:GUV}, it is interesting to note that the UV behavior of the function $\sqrt{pp'}\hat\Pi(p,p')$ assumes a Minkowski-like form with physical momenta in place of time. The Minkowski result $\Pi_{\rm Mink}(t-t',p)$ for the one-loop bubble---in the Minkowski vacuum---is recovered for $p,p'\gg1$ and $|\ln p/p'|=|t-t'|\ll1$:
\beq
\label{appeq:PiUVMink}
 \hat\Pi_{\rm Mink}(\Delta t,p)=\frac{\lambda}{96i\pi^2}\left\{\frac{e^{-ip|\Delta t|}}{|\Delta t|}+\frac{e^{-2i\Lambda |\Delta t|}}{|\Delta t|}\frac{\sin p\Delta t}{p\Delta t}\right\},
\eeq
where we used $\tilde\Lambda|p-p'|\approx\Lambda|t-t'|$.

\section{Influence of subhorizon modes}
\label{appsec:roleofUV}

In this section, we present a complete solution of Eqs. \eqn{eq:Irho}--\eqn{eq:PiH} including high momentum modes. We show that the result \eqn{eq:Irhosol} for the $\rho$ component of the resummed function $\hat I$ is not modified, whereas the $F$ component \eqn{eq:IFsol} is only modified by a constant multiplicative factor of order unity.

The key observation for the present analysis is that the IR and mix behaviors of the one-loop bubble function $\hat \Pi$, Eqs. \eqn{appeq:factorized} and \eqn{appeq:Pimixfac}, have the factorized expressions
\beq
\label{appeq:factoA}
 \hat\Pi_F^{\rm IR}(p,p')=A(p)A(p')
\eeq
and
\beq
\label{appeq:factoB}
 \hat\Pi_F^{\rm mix}(p,p')=A(p)B_F(p')\,,\quad\hat\Pi_\rho^{\rm mix}(p,p')=A(p)B_\rho(p'),
\eeq
where we recall the relation of the one-point function $A(p)$ to the one used in the main text, see \Eqn{eq:Adef},
\beq
 A(p)=i\sqrt{-\pi_F\over p}\,{\cal A}\left(\ln {p\over\mu}\right)
\eeq
and where the functions $B_{F,\rho}(p)$ are obtained from \Eqn{appeq:B}:
\bea
\label{appeq:mombehav1}
 B_F(p)&=&{\rm Re}\,B(p)=-\sqrt{-\pi_F}\frac{\cos p}{p},\\
\label{appeq:mombehav2}
 B_\rho(p)&=&-2{\rm Im}\,B(p)=-2\sqrt{-\pi_F}\frac{\sin p}{p}.
\eea

Let us first consider \Eqn{eq:IF} for IR modes $p,p'\ll1$ and separate the $s$ integral in a low momentum and a high momentum piece: $\int_p^\infty =\int_p^\mu+\int_\mu^\infty$ with $\mu\sim1$. Using the notation \eqn{appeq:IRdef}--\eqn{appeq:ximdef}, we thus write
\bea
\label{appeq:IFIR}
\hat I_F^{{\rm IR}}(p,p')&=&\hat\Pi_H^{{\rm IR}}(p,p')+\int_\mu^\infty ds\,\hat I_\rho^{\rm mix}(p,s)\hat \Pi_H^{{\rm xim}}(s,p')\nn
&+&\int_{p}^\mu ds\,\hat I_\rho^{{\rm IR}}(p,s)\hat \Pi_H^{\rm IR}(s,p').
\eea
The first and third terms on the right-hand side correspond to the IR contribution whereas the second term is the contribution from high momentum modes. The former involves the functions $\hat \Pi_H^{{\rm IR}}$, which itself receives contributions from high momentum modes, as we shall see shortly, and the latter involves the mix behaviors $\hat \Pi_H^{{\rm xim}}$ and $\hat I_\rho^{{\rm mix}}$, which connect IR and UV modes.
Applying a similar treatment to \Eqn{eq:PiH}, we get, for the IR behavior ($p,p'\ll1$),
\bea
\label{appeq:PiHIR}
\hat \Pi_H^{{\rm IR}}(p,p')&=&\hat\Pi_F^{{\rm IR}}(p,p')-\int_\mu^\infty ds\,\hat\Pi_F^{\rm mix}(p,s)\hat I_\rho^{{\rm xim}}(s,p')\nn
&-&\int_{p'}^\mu ds\,\hat\Pi_F^{{\rm IR}}(p,s)\hat I_\rho^{\rm IR}(s,p')
\eea
and, for the ``xim'' behavior ($p'\lesssim\mu\lesssim p$),
\bea
\label{appeq:PiHxim}
\hat \Pi_H^{{\rm xim}}(p,p')&=&\hat\Pi_F^{{\rm xim}}(p,p')-\int_\mu^\infty ds\,\hat\Pi_F^{\rm UV}(p,s)\hat I_\rho^{{\rm xim}}(s,p')\nn
&-&\int_{p'}^\mu ds\,\hat\Pi_F^{{\rm xim}}(p,s)\hat I_\rho^{\rm IR}(s,p').
\eea
Here, the functions $\hat \Pi_H^{{\rm xim}}(p,p')=\hat \Pi_H^{{\rm mix}}(p',p)$ and $\hat\Pi_F^{\rm UV}$ are known whereas the functions $\hat I_\rho^{\rm IR}$ and $\hat I_\rho^{{\rm xim}}$ are to be determined. Applying again the same technique as above to \Eqn{eq:Irho} we get, for the former (which involves $p,p'\lesssim\mu$)
\beq
\label{appeq:IhoIR}
 \hat I_\rho^{\rm IR}(p,p')=\hat\Pi_\rho^{\rm IR}(p,p')+\int_{p}^{p'} ds \,\hat\Pi_\rho^{\rm IR}(p,s )\hat I_\rho^{\rm IR}(s ,p').
\eeq 
The solution of this equation is given in \Eqn{eq:Irhosol} . Indeed, the integral in \Eqn{eq:Irho} for the $\rho$ component only involves momenta $s$ of the order of the external momenta $p$ and $p'$ and thus the IR behavior receives no contribution from UV modes. 

For the function $\hat I_\rho^{{\rm xim}}$, we use the relation $\hat I_\rho^{{\rm xim}}(p,p')=-\hat I_\rho^{{\rm mix}}(p',p)$ and write (so, below, $p\lesssim\mu\lesssim p'$)
\beq
\label{appeq:Ihomix}
 \hat I_\rho^{\rm mix}(p,p')=\hat\Pi_{\rho,H}^{\rm mix}(p,p')+\int_{p}^{\mu} ds \,\hat\Pi_\rho^{\rm IR}(p,s )\hat I_\rho^{\rm mix}(s ,p'),
\eeq
where we defined
\beq
\label{appeq:pirhoH}
 \hat\Pi_{\rho,H}^{\rm mix}(p,p')=\hat\Pi_\rho^{\rm mix}(p,p')+\int_{\mu}^{p'} ds \,\hat\Pi_\rho^{\rm mix}(p,s )\hat I_\rho^{\rm UV}(s ,p').
\eeq
These equations have a similar structure as those we solved when discussing $\hat I_F^{\rm IR}$, i.e., Eqs. \eqn{eq:IF} and  \eqn{eq:PiH}, with the upper bound replaced by $\mu$. We apply a similar technique as that used in Sec. \ref{sec:four} which exploits the fact that the function $\hat\Pi_\rho^{\rm mix}$ factorizes; see \Eqn{appeq:factoB}. In particular, \Eqn{appeq:pirhoH} implies that the function $\hat\Pi_{\rho,H}^{\rm mix}$ can also be factorized:
\beq
 \hat\Pi_{\rho,H}^{\rm mix}(p,p')=A(p)\bar B_\rho(p'),
\eeq
with
\beq
\label{appeq:Bbareq}
 \bar B_\rho(p)=B_\rho(p)-\int_{\mu}^{p} ds\,\hat I_\rho^{\rm UV}(p,s)B_\rho(s).
\eeq
The function $\hat I_\rho^{\rm UV}$ is entirely determined by UV physics. It solves the following equation:
\beq
\label{appeq:IhoUV}
 \hat I_\rho^{\rm UV}(p,p')=\hat \Pi_\rho^{\rm UV}(p,p')+\int_p^{p'}\!\!ds\,\hat \Pi_\rho^{\rm UV}(p,s)\hat I_\rho^{\rm UV}(s,p'),
\eeq
which, for $|\ln p/p'|=|t-t'|\ll1$ reduces to the equation for resumming bubbles in the Minkowski vacuum; see Eqs. \eqn{appeq:PiUV} and \eqn{appeq:PiUVMink}. We note, in particular, that this equation does not generate any infrared enhancement factor. Each new bubble generated by the integral equation thus brings a genuine coupling factor $\lambda$.

Now, using \Eqn{appeq:IhoIR}, one shows that \Eqn{appeq:Ihomix} can be solved as, see Appendix \ref{appsec:proof},
\beq
  \hat I_\rho^{\rm mix}(p,p')=\hat\Pi_{\rho,H}^{\rm mix}(p,p')+\int_{p}^{\mu} ds \,\hat I_\rho^{\rm IR}(p,s )\hat\Pi_{\rho,H}^{\rm mix}(s,p'),
\eeq
from which it follows that
\beq
\label{appsec:Irhomixfac}
  \hat I_\rho^{\rm mix}(p,p')=\bar A(p)\bar B_\rho(p'),
\eeq
with
\beq
\label{appeq:Abar}
 \bar A(p)=A(p)+\int_{p}^{\mu} ds\,\hat I_\rho^{\rm IR}(p,s) A(s)\,.
\eeq
We note that \Eqn{appeq:Abar} is identical to \Eqn{eq:abareq} which we solved in Sec. \ref{sec:four}. We thus have
\beq
 \bar A(p)=i\sqrt{-\pi_F\over p}\,\bar {\cal A}\left(\ln {p\over\mu}\right)=\frac{i\sqrt{-\pi_F}}{p^{\bar\kappa+{1\over2}}},
\eeq
with $\bar\kappa=\bar\nu-\varepsilon$.

Inserting $\hat I_\rho^{\rm xim}(p,p')=-\hat I_\rho^{\rm mix}(p',p)=-\bar B_\rho(p)\bar A(p')$ in \eqn{appeq:PiHIR}, we obtain
\beq
\label{appeq:PiHIRfinal}
 \hat \Pi_H^{{\rm IR}}(p,p')=A(p)\bar A(p')\left\{1+\int_\mu^\infty ds\,\bar B_\rho(s)B_F(s)\right\},
\eeq
where we used the factorization property \eqn{appeq:factoA} as well as \Eqn{appeq:Abar}. Similarly, we get, from \eqn{appeq:PiHxim},
\beq
\label{appeq:PiHximfinal}
 \hat \Pi_H^{{\rm xim}}(p,p')=\left\{B_F(p)+\int_\mu^\infty ds\,\hat \Pi^{\rm UV}_F(p,s)\bar B_\rho(s)\right\}\bar A(p').
\eeq
We emphasize that the integrals in both \eqn{appeq:PiHIRfinal} and \eqn{appeq:PiHximfinal} are finite in $d=3$ as can be checked from Eqs. \eqn{appeq:PiUV}, \eqn{appeq:mombehav1}, \eqn{appeq:mombehav2} and \eqn{appeq:Bbareq}.
All the ingredients entering the right-hand side of \Eqn{appeq:IFIR} have now been determined. Using again \Eqn{appeq:IhoIR}, we finally obtain
\beq
\label{appeq:IFfinal}
 \hat I_F^{\rm IR}(p,p')=Z\bar A(p)\bar A(p'),
\eeq
with the renormalization factor
\bea
\label{appeq:Zfactor}
 Z&=&1+2\int_\mu^\infty ds\,\bar B_\rho(s)B_F(s)\nn
 &+&\int_\mu^\infty dsds'\bar B_\rho(s)\hat\Pi_F^{\rm UV}(s,s')\bar B_\rho(s').
\eea

It is remarkable that  \Eqn{appeq:IFfinal}, which fully takes into account UV modes, is essentially identical to the solution obtained in Sec. \ref{sec:four}, see Eqs. \eqn{eq:IFsol} and \eqn{eq:IFresult}, where UV modes were neglected. Superhorizon modes only affect the final result through a constant, finite renormalization factor $Z$. This important result shows that there is an effective decoupling of UV and IR physics in de Sitter space.

The fact that the statistical two-point function $\hat I_F$ (the two-point statistical function of the composite operator $\varphi^2$) takes a factorized form, \Eqn{appeq:IFfinal}, in the IR is a remarkable result too. As already emphasized in Sec. \ref{sec:four}, this is reminiscent of the fact that de Sitter IR fluctuations can be described by a classical stochastic ensemble of growing modes \cite{Guth:1985ya,Polarski:1995jg,Campo:2003pa}. Here, we emphasize another interesting result, namely the fact that the spectral component $\hat I_\rho$ also factorizes in the mix regime, see \Eqn{appsec:Irhomixfac}. This is also rooted in the classical statistical nature of IR fluctuations in de Sitter and is to be put in parallel with the corresponding factorized form of the spectral function of the field $\varphi$, \Eqn{appeq:rhomix}.

We end this section with a few remarks concerning the renormalization factor \eqn{appeq:Zfactor}. The parametric dependence of the various factors in \eqn{appeq:Zfactor} on the coupling constant $\lambda$ is as follows. From Eqs. \eqn{appeq:PiUV} and \eqn{appeq:IhoUV} we get $\hat \Pi^{UV}\sim\lambda$ and $\hat I^{UV}\sim\lambda(1+{\cal O}(\lambda))$. This implies that $\bar B_{\rho}\sim B_{\rho}(1+{\cal O}(\lambda))$, see \Eqn{appeq:Bbareq}. Recalling that $B_{F,\rho}\sim\sqrt{\pi_\rho}$, where $\pi_\rho\sim\lambda/\varepsilon$, Eqs. \eqn{appeq:mombehav1} and \eqn{appeq:mombehav2}, we thus have
\beq
 Z-1\sim\pi_\rho(1+{\cal O}(\lambda)).
\eeq
As discussed in Sec. \ref{sec:discussion}, one has either $\pi_\rho\sim\sqrt\lambda$, in the case of a vanishing tree-level square mass, or $\pi_\rho\sim 1$ in the case of negative tree-level square mass. In any case, for small coupling we can approximate $\bar B_{F,\rho}\approx B_{F,\rho}$ and write, using \eqn{appeq:mombehav1} and \eqn{appeq:mombehav2},
\bea
 Z&\approx&1-2\pi_F\int_\mu^\infty ds\, \frac{\sin 2s}{s^2}\nn
 &\approx&1+2\pi_\rho\left\{{\sin 2\mu\over2\mu}-{\rm Ci}(2\mu)\right\},
\eea
where we used $\pi_F=-\pi_\rho F_\nu\approx-\pi_\rho/2$ in $d=3$.

\section{Renormalization}
\label{appsec:renorm}

The present theory is perturbatively renormalizable in $D=d+1=4$. The renormalization of the gap equation \eqn{eq:mass} has been discussed in \cite{Serreau:2011fu}. The main result is that it can be made finite by a standard (Minkowski-like, i.e., curvature independent) redefinition of the parameters $m^2$, $\xi$ and $\lambda$. Here, we discuss (for the first time to our knowledge in the present context) the renormalization of the four-point function \Eqn{eq:gamma4}, or \eqn{eq:fourpointconf}.

As in previous sections, we regularize the theory by a simple sharp cutoff on physical momenta, i.e. by using the regularized propagator $\hat G(p,p')\to\hat G(p,p')\theta(\Lambda-p)\theta(\Lambda-p')$. Such a cutoff on physical momenta actually breaks the de Sitter symmetry but is compatible with the $p$-representation, i.e., it does not break the affine subgroup of the de Sitter group, which is at the root of the $p$-representation \cite{Busch:2012ne}. Moreover, the local Lorentz---and thus de Sitter--- symmetry is restored after renormalization provided one imposes suitable, i.e., covariant renormalization conditions. An alternative would be to employ de Sitter invariant regulators, such as dimensional \cite{Weinberg:2005vy,Seery:2007we,Senatore:2009cf}, or Pauli-Villars \cite{Weinberg:2010wq} regularizations. 

When discussing renormalization, i.e., the removal of the regulator, the decomposition \eqn{eq:Pidec} should be modified to take into account a local term
\beq
 \hat\Pi(p,p')=i\hat\pi\delta_{\hat\C}(p-p')+\hat\Pi_F(p,p')-{i\over2}{\rm sign}_{\hat \C}(p-p')\hat\Pi_\rho(p,p'),
\eeq
where $\hat\pi$ is a (divergent) constant. This expresses the fact that $i\hat\Pi(p,p')$ is a one-loop contribution to the four-point function, see \eqn{eq:fourpointconf}, and thus potentially contains a (divergent) contribution with the same structure as the tree-level (bare) coupling\footnote{Equivalently, $i\hat\Pi(p,p')$ is the one-loop contribution to the two-point correlator of the composite field $\varphi^2$, whose tree-level contribution is $\sim\delta_{\hat{\C}}(p-p')$; see \eqn{eq:Dprep}.} $\sim\delta_{\hat\C}(p-p')$; see \eqn{eq:Dprep}. Let us now give a general analysis of the divergent contribution.

We recall the expression with $d=3$ spatial dimensions
\beq
 \hat\Pi(p,p')=-{\lambda\over6}\int_\bq\frac{\hat G(qp,qp')\hat G(rp,rp')}{qr},
\eeq
with $r=|\be+\bq|$. The divergent contribution $\hat\Pi^{\rm div}$ comes from very high momentum modes $1\ll\mu_r <qp,qp'< \Lambda$, where $\mu_r$ is an arbitrary (renormalization) scale. Specializing to the case $p, p'\in\hat\C^+$ for simplicity,\footnote{The general result has the same structure with the replacement $|p-p'|\to-(p-p'){\rm sign}_{\hat{\C}}(p-p')$.} we thus have\footnote{Note that it is sufficient to approximate $r\approx q$ in order to get the leading (logarithmic) divergent behavior.}
\bea
 \hat\Pi^{\rm div}(p,p')&=&-{\lambda\over6}\int_{\tilde\mu_r<|\bq|<\tilde\Lambda}\frac{\hat G^2_{\rm UV}(qp,qp')}{q^2}\\
 &=&-{\lambda\over48\pi^2}\int_{\tilde\mu_r}^{\tilde\Lambda}dq\,e^{-2iq|p-p'|}\\
 &=&{\lambda\over96\pi^2}\frac{e^{-2i\tilde\Lambda|p-p'|}-e^{-2i\tilde\mu_r|p-p'|}}{i|p-p'|},
\eea
where $\tilde\Lambda\equiv\Lambda/\max(p,p')$ and $\tilde\mu_r\equiv\mu_r/\max(p,p')$. This expression agrees with our previous findings Eqs. \eqn{appeq:UVcontrib2} and \eqn{appeq:PiUV}. 

To extract the divergent contribution in the limit $\Lambda\to\infty$, we use the well-known result
\beq
\label{appeq:divsin}
  \frac{\sin 2\tilde\Lambda |x|}{|x|}\to\pi\delta(x)
\eeq
as well as 
\beq
\label{appeq:divcos}
 \frac{\cos2\tilde\Lambda|x|}{|x|}\to2\ln{\mu_r\over\Lambda}\delta(x)+\frac{\cos2\tilde\mu_r|x|}{|x|}.
\eeq
The latter can be shown as follows. For any test function $f(x)$, one has
\bea
 &&\int dx\,f(x)\frac{\cos2\tilde\Lambda|x|-\cos2\tilde\mu_r|x|}{|x|}\nn
 &&\to2f(0)\int_0^\infty\!\!dy\,\frac{\cos y-\cos{\mu_r\over\Lambda}y}{y}\\
 &&=2f(0)\lim_{\eta\to0}\left[{\rm Ci}\left({\mu_r\eta\over\Lambda}\right)-{\rm Ci}(\eta)\right]\\
 &&=2f(0)\ln{\mu_r\over\Lambda},
\eea
where we used $\tilde\mu_r/\tilde\Lambda=\mu_r/\Lambda$. We made the change of variable $y=2\Lambda x$ and took the limit $\Lambda/\mu_r\gg1$ in the second line. The third line is obtained by noticing that, since the integrand vanishes at $y=0$, one can safely replace $\int_0^\infty dy=\lim_{\eta\to0}\int_\eta^\infty$. Finally, we used
${\rm Ci}(x)=-\int_x^\infty dt\,\cos t/t=\ln x+\gamma +{\cal O}(x^2)$ at small $x$.
We thus get the desired divergent contribution as
\beq
\label{appeq:divergence}
 \hat\Pi^{\rm div}(p,p')\to\frac{i\lambda}{48\pi^2}\ln{\Lambda\over\mu_r}\,\delta(p-p').
\eeq
We note that, since the UV divergent term is local, one can replace $p-p'\to-p(t-t')$. One recovers the usual Minkowski one-loop coupling divergence for $\sqrt{pp'}\hat\Pi(p,p')\to\Pi_{\rm Mink}(t-t',p)$:
\beq
 \Pi^{\rm div}_{\rm Mink}(t-t',p)\to\frac{i\lambda}{48\pi^2}\ln{\Lambda\over\mu_r}\,\delta(t-t').
\eeq
This guarantees that the divergence \eqn{appeq:divergence} can be absorbed by the usual Minkowski counterterm. Incidentally, this ensures that the local Lorentz---hence de Sitter---symmetry is properly restored in the usual way after the UV cutoff is removed.

We are now in a position to discuss the renormalization of the resummation equation \eqn{eq:Np2}. For this purpose, it is convenient to extract the explicit coupling dependence of each term. We define
\beq
\label{appeq:couplings}
 \hat\Pi(p,p')=\lambda\pi(p,p')\,,\quad \hat I(p,p')=\lambda^{-1} \Im(p,p'),
\eeq
such that $\pi\sim\lambda^0$ and $\Im\sim\lambda^2$. Equation \eqn{eq:Np2} thus reads
\beq
\label{appeq:Ieqeq}
 \Im (p,p')=\lambda^2\pi(p,p')-i\lambda\int_{\hat\C} ds\,\pi(p,s)\Im(s,p').
\eeq
We have, from the above discussion,
\beq
\label{appeq:pidivdef}
 \pi(p,p')=i\pi_{\rm div}\delta_{\hat\C}(p-p')+\pi^f(p,p'),
\eeq
where $\pi^f(p,p')$ is finite and
\beq
 \pi_{\rm div}=-\frac{1}{48\pi^2}\ln{\Lambda\over\mu_r}.
\eeq
Accordingly, we write
\beq
\label{appeq:Idivdef}
 \Im(p,p')=i\Im_{\rm div}\delta_{\hat\C}(p-p')+\Im^f(p,p'),
\eeq
where $\Im^f(p,p')$ is finite, as shown below. Plugging the expressions \eqn{appeq:pidivdef} and \eqn{appeq:Idivdef} in \Eqn{appeq:Ieqeq}, we get
\beq
\label{appeq:Idiv}
 \Im_{\rm div}=\frac{\lambda^2\pi_{\rm div}}{1-\lambda\pi_{\rm div}}
\eeq
and, after some simple manipulations,
\beq
\label{appeq:finiteeqI}
 \Im^f(p,p')=\lambda_r^2\pi^f(p,p')-i\lambda_r\int_{\hat\C}ds\,\pi^f(p,s)\Im^f(s,p'),
\eeq
where we introduced the renormalized coupling
\beq
\label{appeq:renormcoupling}
 \lambda_r=\frac{\lambda}{1-\lambda\pi_{\rm div}}=\frac{\lambda}{1+\frac{\lambda}{48\pi^2}\ln{\Lambda\over\mu_r}}.
\eeq
This is the standard expression for the renormalized coupling in the large-$N$ limit, up to possible (renormalization-scheme-dependent) finite parts \cite{Coleman:1974jh}. It shows the trivial character of the present theory, i.e., that for any positive bare coupling $\lambda$, the running coupling $\lambda_r\to 0^+$ in the IR, i.e., for $\mu_r/\Lambda\to0$. Reversely, demanding that $\lambda_r$ be finite, \Eqn{appeq:renormcoupling} expresses how the bare coupling $\lambda$ evolves as the cutoff is removed: ${1/\lambda}={1/\lambda_r}+\pi_{\rm div}$. One get the usual Landau pole in the UV, the value $\Lambda_L$ of the cutoff where the bare coupling diverges: $\Lambda_L/\mu_r=\exp(48\pi^2/\lambda_r)$. Finally, we stress that the renormalized coupling \eqn{appeq:renormcoupling} coincides with that needed to absorb coupling subdivergences in the equation for self-energy $\Sigma$ \cite{Serreau:2011fu}, as expected.
 
Using the large momentum behavior of the $F$ and $\rho$ components of the function $\pi^f(p,s\to\infty)$, which can be inferred from the previous sections, see Eqs. \eqn{appeq:Pimixfac}, \eqn{appeq:B} and \eqn{appeq:PiUV}, one can show that the contour integral in \eqn{appeq:finiteeqI} does not lead to UV divergence. This is related to the fact that this is originally a time integral, and not a momentum loop integral. Thus, we conclude that the function $\Im^f(p,p')$ is finite when expressed in terms of the renormalized coupling $\lambda_r$. To make link with the original functions $\hat\Pi$ and $\hat I$ discussed in the paper, we define the finite functions 
\beq
 \hat\Pi^r(p,p')=\lambda_r\pi^f(p,p')\,,\quad\hat I^r(p,p')=\lambda_r^{-1}\Im^f(p,p'),
\eeq
in terms of which \Eqn{appeq:finiteeqI} reads
\beq
\label{appeq:renqties}
 \hat I^r(p,p')=\hat\Pi^r(p,p')-i\int_{\hat\C} ds \,\hat\Pi^r(p,s )\hat I^r(s ,p').
\eeq
This is to be compared to \Eqn{eq:Np2}. This shows that the discussion presented in the main body of the paper (including appendices) readily applies to the renormalized quantities defined here provided one replaces bare quantities $\lambda$, $\hat\Pi$, and $\hat I$ by their renormalized counterparts $\lambda_r$, $\hat\Pi^r$, and $\hat I^r$, together with the replacement $\Lambda\to\mu_r$ in momentum loop integrals.

To close this section, we note that the bare function $\hat I(p,p')$ is not finite, but contains a local divergence $\sim i\lambda_r\pi_{\rm div}\delta_{\hat{\C}}(p-p')$, see Eqs. \eqn{appeq:couplings},\eqn{appeq:Idivdef}, and \eqn{appeq:Idiv}. This is, however, not a problem since $\hat I$ is not directly a physical quantity but merely the loop correction to either the two-point correlator $\hat D$, see \Eqn{eq:Dprep}, or, equivalently, the four-point vertex function, see \Eqn{eq:gamma4}, or \eqn{eq:fourpointconf}. One easily checks that these physical quantities are finite after renormalization. Using Eqs. \eqn{appeq:couplings},\eqn{appeq:Idivdef}, \eqn{appeq:Idiv}, \eqn{appeq:renormcoupling}, and \eqn{appeq:renqties}, one finds that
\bea
 i\hat D(p,p')&=&-\frac{\lambda}{3N}\left[-\delta_{\hat\C}(p-p')+i\hat I(p,p')\right]\nn
 &=&-\frac{\lambda_r}{3N}\left[-\delta_{\hat\C}(p-p')+i\hat I^r(p,p')\right]
\eea
is finite and thus so is the four-point vertex function \eqn{eq:fourpointconf}.

\end{document}